\documentclass[cernpreprint,USenglish]{na61doc}
\usepackage{amsmath}
\usepackage{url}
\usepackage{lscape}
\usepackage{multirow}
\usepackage{makecell}

\usepackage{colortbl}
\definecolor{darkred}{rgb}{0.5,0,0}
\definecolor{darkblue}{rgb}{0,0,0.5}
\definecolor{firebrick}{rgb}{0.75,0.125,0.125}
\definecolor{darkgreen}{rgb}{0,0.5,0}

\usepackage[colorlinks=true,linkcolor=firebrick,citecolor=darkgreen,urlcolor=darkblue]{hyperref}

\usepackage{xspace}
\usepackage{enumerate}

\ShineTitle{Measurements of higher-order cumulants of multiplicity and net-electric charge distributions
         in inelastic proton-proton interactions by \NASixtyOne} 

\PreprintIdNumber{CERN-EP-2023-290}

\ShineJournal{}

\ShineJournalRef{}
\ShineDOI{}

\ShineAbstract{
This paper presents the energy dependence of multiplicity and net-electric charge fluctuations in \pp interactions at beam momenta 20, 31, 40, 80, and 158 \GeVc. Results are corrected for the experimental biases and quantified with the use of cumulants and factorial cumulants. 
Cumulant ratios are an essential tool in the search for the critical point of strongly interacting matter in heavy ion collisions. Measurements performed in \pp interactions provide a vital baseline estimation in these studies. The measured signals are compared with the string hadronic models \EposLong and FTFP-BERT. 
}

\newcommand{\eV}{\ensuremath{\mbox{e\kern-0.1em V}}\xspace}
\newcommand{\GeV}{\ensuremath{\mbox{Ge\kern-0.1em V}}\xspace}
\newcommand{\MeV}{\ensuremath{\mbox{Me\kern-0.1em V}}\xspace}
\newcommand{\GeVc}{\ensuremath{\mbox{Ge\kern-0.1em V}\!/\!c}\xspace}
\newcommand{\GeVcc}{\ensuremath{\mbox{Ge\kern-0.1em V}\!/\!c^2}\xspace}
\newcommand{\AGeV}{\ensuremath{A\,\mbox{Ge\kern-0.1em V}}\xspace}
\newcommand{\AGeVc}{\ensuremath{A\,\mbox{Ge\kern-0.1em V}\!/\!c}\xspace}
\newcommand{\MeVc}{\ensuremath{\mbox{Me\kern-0.1em V}/c}\xspace}

\newcommand{\cm}{\ensuremath{\mbox{cm}}\xspace}

\newcommand{\dd}{\ensuremath{{\text{d}}}\xspace}
\newcommand{\dedx}{\ensuremath{\dd E\!/\!\dd x}\xspace}

\newcommand{\pp}{\mbox{\textit{p+p}}\xspace}

\newcommand{\coordinate}[1]{{\fontfamily{lmss}\selectfont#1}}

\newcommand{\EposLong}{{\scshape Epos1.99}\xspace}

\newcommand{\CernVM}{\textsc{Cern\-\kern-0.05emVM}\xspace}

\begin{document}

\maketitle
\section{Introduction}

This paper presents experimental results on event-by-event
fluctuations of multiplicities of all charged, positively, and negatively charged hadrons as well as fluctuations of net-electric charge (so-called net-charge) produced in inelastic proton-proton (\pp) interactions
at beam momenta 20, 31, 40, 80, and 158~\GeVc. The corresponding energy per nucleon pair in the center-of-mass system is 6.3, 7.7, 8.8, 12.3, and 17.3~\GeV, respectively.

The measurements were performed by the multi-purpose
\mbox{\NASixtyOne}~\cite{Abgrall:2014xwa} spectrometer at
the CERN Super Proton Synchrotron (SPS) in 2009 as a part of the strong interaction program. This program is devoted to studying the
properties of the onset of deconfinement and searching for the critical point~(CP) of the strongly interacting matter~\cite{Aduszkiewicz:2642286}.
 The phase diagram is usually represented by temperature versus baryo-chemical potential. At baryo-chemical potential $\mu_{B}\approx 0$, the transition from hadronic phase to quarks and gluons is a rapid but smooth cross-over as predicted by lattice calculations~\cite{Aoki:2006br,Borsanyi:2010bp, Borsanyi:2015waa}. 
At low temperature $T$ and large $\mu_{B}$, the phenomenological models predict the first-order phase transition~\cite{Stephanov:2004xs} which ends with the second-order critical endpoint. 

Fluctuation and correlation analyses may be sensitive to CP~\cite{Stephanov:1999zu, Shuryak:2000pd} due to their connection with correlation length. Other effects may dilute the CP signal, e.g., 
local and global conservation laws~\cite{BRAUNMUNZINGER2021122141}. Measurements performed in \pp interactions provide a vital baseline estimation which may be crucial to understand signal measured in heavy ion collisions. 
Results on \pp interactions also give a unique opportunity to test models of strong interactions which help to understand results on nucleus-nucleus collisions. Rich experimental data in \pp interactions on particle production in full phase space is already available from bubble-chamber or streamer experiments~\cite{Heiselberg:2000fk}. It should be underlined that the data statistics of these experiments are considerably smaller than nowadays measurements (see Sec.~\ref{sec:analysis}). 
In case of fluctuations, those measurements and predictions (like KNO-G scaling~\cite{Koba:1972ng, Golokhvastov:2001ei, Golokhvastov:2001pt}) are difficult to be compared to modern experimental analysis due to different analysis acceptance~\cite{Savchuk:2019xfg}. 

Throughout this paper, the rapidity: $y=0.5\ln[(E+cp_\text{L})/(E-cp_\text{L})]$, is calculated in the
collision center-of-mass system by shifting rapidity in the laboratory frame by rapidity of the center-of-mass, assuming proton mass. The $E$, $p_\text{L}$, and $c$ are the particle energy (assuming pion mass for a given charged particle), its longitudinal momentum, and the velocity of light, respectively.
The transverse component of the momentum is denoted as $p_\text{T}$, and 
the azimuthal angle $\phi$ is the angle between the transverse momentum vector and the horizontal
(\coordinate{x}) axis. Total momentum in the laboratory system is denoted as $p$. 
The collision energy per
nucleon pair in the center-of-mass system is denoted as $\sqrt{s_{\emph{NN}}}$.
\section{Intensive quantities}
\label{sec:IQ}
Net-charge, as well as multiplicity fluctuations, are one of the tools to search for CP in nucleus-nucleus collisions.
Although there is great freedom in selecting fluctuation measures, it is reasonable to choose ones sensitive to the desired physical phenomenon and insensitive to other possible sources of fluctuations, e.g., \textit{system volume}~($V$). As \pp interactions are measured as a reference for nucleus-nucleus collisions, choosing quantities that make comparing systems easier is particularly important. 

It is more convenient to present results 
in terms of cumulants than moments of the multiplicity distribution~\cite{Asakawa:2015ybt}. Cumulants and moments are proportional to $V$ and called extensive variables ($\sim V$)~\cite{Asakawa:2015ybt}.
If the event quantity $N$ is measured, then the $n$-th order moment of its probability distribution, $P(N)$, is defined by
	\begin{equation}
		\langle N^{n} \rangle = \sum_{N} N^{n}P(N),
	\end{equation}
 where $\langle\dots\rangle$ denotes averaging over events.
Then, the first four cumulants ($\kappa_{n}$, $n\leq4$) are given by
\begin{eqnarray}
	\kappa_{1}[N]=\langle N\rangle, \\
	\kappa_{2}[N]=\langle \delta N^{2}\rangle, \\
	\kappa_{3}[N]=\langle \delta N^{3}\rangle, 	\\
	\kappa_{4}[N]=\langle \delta N^{4}\rangle-3\langle\delta N^{2}\rangle^{2},
\end{eqnarray}
where $\langle\delta N^{n} \rangle=\langle(N-\langle N \rangle)^{n}\rangle$. 


A ratio of two extensive quantities is an intensive quantity, e.g., the scaled variance of $N$:
		\begin{equation}
			\omega[N] = \frac{\kappa_{2}[N]}{\kappa_{1}[N]}.
		\end{equation}
		
The scaled variance calculated within a simple model like the ideal Boltzmann gas described in the Grand Canonical Ensemble (GCE) reads~\cite{Gorenstein:2011vq}:
\begin{equation}
\omega[N] = \omega[N]^{*} + \langle N \rangle / \langle V \rangle \cdot \omega[V], \\ 
\label{eq:gce:varN}
\end{equation}
where $\omega[N]^{*}$ stands for the scaled variance at fixed volume $V$. 
The first component $\omega[N]^{*}$ of Eq. 7 is considered the \textit{wanted} one, and it is \emph{independent of the volume fluctuations}. However, the second component is seen as \textit{unwanted} as it is trivially proportional to the scaled variance of the volume distribution $\omega[V]$. 

		
In GCE, intensive quantity $\omega$ has the following features:
		\begin{enumerate}[(i)]
			\item it is independent of $V$ (for event ensembles with fixed $V$),
			\item it depends on fluctuations of $V$ (even if $\langle V \rangle$ is fixed),
			\item for Poisson distribution it is equal to unity.
		\end{enumerate}
	
For third and fourth-order cumulants, one can construct intensive quantities similarly as:
\begin{equation}
\frac{\kappa_{3}[N]}{\kappa_{2}[N]}, 
\quad \frac{\kappa_{4}[N]}{\kappa_{2}[N]}. 
\label{Eq:HM}
\end{equation}
These quantities also are intensive so independent of volume but they remain sensitive to the $V$ \emph{fluctuations}~\cite{Bialas:1976ed,Gorenstein:2011vq,Begun:2016sop}. 

The Poisson distribution is considered the reference as particles produced in GCE will follow it with the $\lambda$ parameter being equal to the average multiplicity of a given particle type~\cite{Gorenstein:2008nr, Begun:2006jf}. The sum of charges in the ideal gas model also will follow the Poisson distribution. 
In the ideal gas model, the net-charge distribution will be the Skellam distribution, which is defined as a difference between two independent Poisson distributions of positively and negatively charged particles with constants $\langle h^{+}\rangle$ and $\langle h^{-}\rangle$, where $h^{+}$ and $h^{-}$ stand for multiplicities of positively and negatively charged hadrons. 
 The following relation gives the Skellam distribution cumulants: $\kappa_{i}=\langle h^{+}\rangle + (-1)^{i}\langle h^{-}\rangle$, where $i$ is the cumulant order. 
In such a case, ratios of even and odd cumulants will not keep one as a reference value.
The modification of the reference for net-charge, so it remains one and is intensive can be introduced in the following way:
\begin{equation}
\frac{\kappa_{2}[h^{+}-h^{-}]}{\kappa_{1}[h^{+}]+\kappa_{1}[h^{-}]},  \quad  \frac{\kappa_{3}[h^{+}-h^{-}]}{\kappa_{1}[h^{+}-h^{-}]}.      
\end{equation}


The use of intensive quantities is crucial in the case of comparisons between \pp and nucleus-nucleus collisions. Cumulants mix information about correlations of different orders; for details of the relation see Ref.~\cite{Bzdak:2019pkr}. Factorial cumulants ($\hat{C}_{n}$, $n\geq2$) are constructed to cancel all lower-order correlations so only a given order of correlations can be studied~\cite{Bzdak:2019pkr, Ling:2015yau, Bzdak:2016sxg}. 
The factorial cumulants are defined for a single-charge case as:
\begin{align}
    \hat{C}_{2}[N] =& -\kappa_{1}[N]+\kappa_{2}[N],\\
    \hat{C}_{3}[N] =& 2\kappa_{1}[N]-3\kappa_{2}[N]+\kappa_{3}[N],\\
    \hat{C}_{4}[N] =& -6\kappa_{1}[N]+11\kappa_{2}[N]-6\kappa_{3}[N]+\kappa_{4}[N].
\end{align}
The Poisson baseline for factorial cumulants ($n\geq2$) is zero.

\section{Experimental setup}\label{sec:setup}

\begin{figure*}
			\centering
			\includegraphics[width=0.8\textwidth]{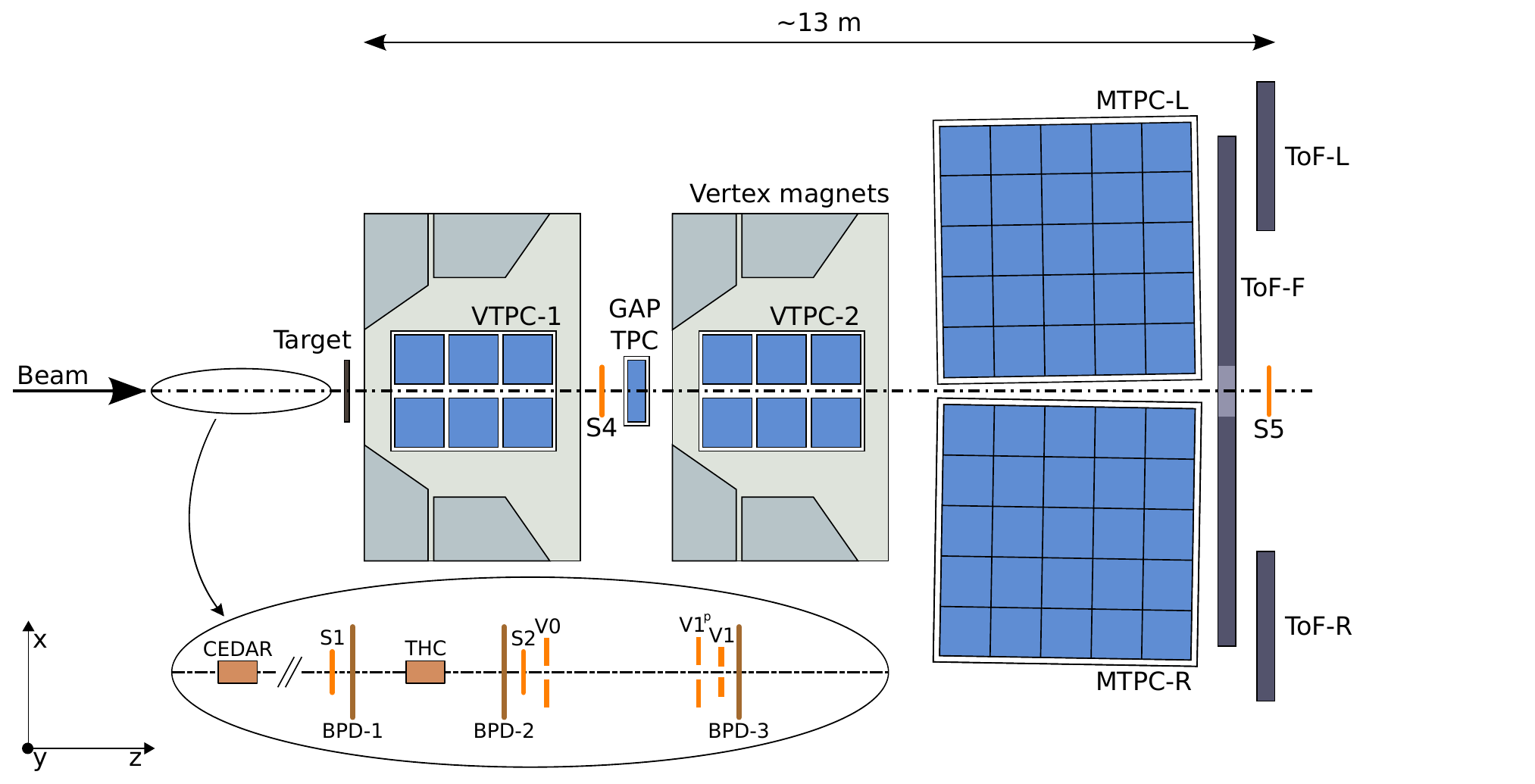}
			\caption[]{
				(Color online) The schematic layout of
				the \NASixtyOne experiment during \pp data taking in 2009 at the CERN SPS
				(horizontal cut, not
				to scale), see text and Ref.~\cite{Abgrall:2014xwa} for details.
			}
			\label{fig:detector-setup}
		\end{figure*}

The \NASixtyOne experiment~\cite{Abgrall:2014xwa} is a large 
acceptance hadron spectrometer located in the H2 beam line of the CERN North Area. 
The schematic layout of the \NASixtyOne detector components is shown in Fig.~\ref{fig:detector-setup}. 

The results presented in this paper were obtained using measurements from
the Time Projection Chambers (TPC), the Beam Position Detectors (BPD), and the beam 
and trigger counters. 
As many publications concerning \pp data-taking in the \NASixtyOne are available, we provide only a brief description of the detector system. The detector elements, the proton beam, the liquid hydrogen target, and the data reconstruction procedure are
described in detail in Refs.~\mbox{\cite{Abgrall:2014xwa, NA61SHINE:2013tiv, NA61SHINE:2017fne}}.

For data taking on \pp interactions, a liquid hydrogen target of 20.29~cm
length (2.8\%~interaction length) and 3~cm diameter was placed 88.4~cm 
upstream of VTPC-1.

Secondary beams of positively charged hadrons at 20, 31, 40, 80, and
158~\GeVc were produced from 400~\GeVc
protons extracted from the SPS onto a beryllium target. 
Protons from the secondary hadron beam 
were identified by two Cherenkov
counters, a CEDAR (either \mbox{CEDAR-W} or \mbox{CEDAR-N}) and a threshold counter 
(THC). 
Due to their limited range of operation, two different
CEDAR counters were employed,
namely for beams at 31, and 40~\GeVc the \mbox{CEDAR-W} counter and
for beams at 80 and 158~\GeVc the \mbox{CEDAR-N} counter. The threshold counter
was used for all beam energies.
A selection based on signals from the Cherenkov counters allowed the identification of
beam protons with a purity of
about 99\%. A coincidence of these signals provided the beam trigger $T_\text{beam}$.


A set of scintillation, Cherenkov counters, and beam
position detectors (BPDs) upstream of the spectrometer provide
timing reference, identification, and position measurements
of incoming beam particles. Trajectories of individual beam
particles were measured in a telescope of beam position detectors
placed along the beamline (BPD-1/2/3 in Fig.~\ref{fig:detector-setup}).
 
Two scintillation counters, S1 and S2, provided beam definition, together
with the three veto
counters V0, V1 and V1$^\mathrm{p}$ with a 1~cm diameter hole. 
The S1 counter also provided the timing (start time for the gating of all counters).
Beam protons were then selected by the coincidence:
\begin{equation}
T_\text{beam} =
\textrm{S1}
\wedge\textrm{S2}
\wedge\overline{\textrm{V0}}
\wedge\overline{\textrm{V1}}
\wedge\overline{\textrm{V1}^\mathrm{p}}
\wedge\textrm{CEDAR}
\wedge\overline{\textrm{THC}}~.
\end{equation}
The interaction trigger $T_\text{int}$ was provided by the anti-coincidence of
the incoming hadron beam and a scintillation counter S4
($T_\text{int} = T_\text{beam} \wedge\overline{\textrm{S4}}$).
The S4 counter, with a two-centimeter diameter, was placed
between the VTPC-1 and VTPC-2 detectors along the beam trajectory
at about 3.7~m from the target,
see Fig.~\ref{fig:detector-setup}.

The main tracking devices of the spectrometer are four
large volume TPCs. Two of them,
the vertex TPCs (\mbox{VTPC-1} and \mbox{VTPC-2}), are
located in the magnetic fields of two super-conducting dipole
magnets with a maximum combined bending power of 9~Tm, which
corresponds to about 1.5~T and 1.1~T fields in the upstream
and downstream magnets, respectively. 
To optimize
the acceptance of the detector, the fields in both magnets were
set in proportion to the beam momentum.
Two large main TPCs (\mbox{MTPC-L} and \mbox{MTPC-R})
are positioned downstream of the magnets symmetrically to
the beam line. The fifth small TPC (GAP TPC) is placed
between \mbox{VTPC-1} and \mbox{VTPC-2} directly on the
beam line. 
The TPCs are filled with Ar:CO$_2$ gas mixtures in proportion 90:10 for
the VTPCs and the GAP TPC, and 95:5 for the MTPCs.
The TPCs provide measurements of energy loss \dedx of charged particles in the chamber gas
along their trajectories. 
Simultaneous measurements of \dedx and $p$ allow extracting information on
particle mass, which is used to identify charged particles. In the case of this analysis, \dedx is used only for electron contamination removal.

\section{Analysis}
\label{sec:analysis}
This section starts with a brief overview of the data analysis procedure and the applied corrections.
It also defines which class of particles the final results correspond to.

The final results refer to charged hadrons 
produced in inelastic \pp interactions 
by strong interaction processes and electromagnetic
decays of produced hadrons. 
Such hadrons are referred to as \emph{primary} hadrons obtained within the analysis acceptance~\cite{na61ArScAcc} (for details, see Sec.\ref{sec:acceptance}). Considered charge combinations are indicated as:
\begin{enumerate}[(i)]
    \item $h^{+} + h^{-}$ -- all charged hadrons,
    \item $h^{+}$ -- positively charged hadrons,
    \item $h^{-}$ -- negatively charged hadrons,
    \item $h^{+} - h^{-}$ -- net-charge being defined as the difference between positively and negatively charged hadrons in a given event.
\end{enumerate}
The availability of the whole distributions and their cumulants $\kappa_{i}$ should allow the reader to obtain the desired quantity if it is not provided here.

The analysis procedure consists of the following steps:
\begin{enumerate}[(i)]
	\item application of event and track selection criteria,
	\item determination of all charged, positively, and negatively charged hadron multiplicity distributions as well as the net-charge distributions,
	\item evaluation of corrections to the distributions based on
	experimental data and simulations,
	\item calculation of the corrected moments and fluctuation quantities,
	\item calculation of statistical and estimation of systematic uncertainties.
\end{enumerate}

Corrections for the following biases were evaluated and applied:
\begin{enumerate}[(i)]
	\item contribution of particles other than \emph{primary} 
	hadrons produced in inelastic \pp interactions,
	\item losses of \emph{primary} hadrons due to measurement inefficiencies, 
	\item losses of inelastic \pp interactions due to the trigger and the
	event and track selection criteria employed in the analysis.
\end{enumerate}
The corrections are calculated using the unfolding procedure~\cite{unfolding} performed on the distributions of the given charge combination after the event and track selection. 
The analysis acceptance was taken from Ar+Sc analysis~\cite{na61ArScAcc} (see Sec.~\ref{sec:acceptance}). 

\subsection{Analysis acceptance}
\label{sec:acceptance}
\begin{figure}[h]
			\centering
			\includegraphics[width=0.4\textwidth]{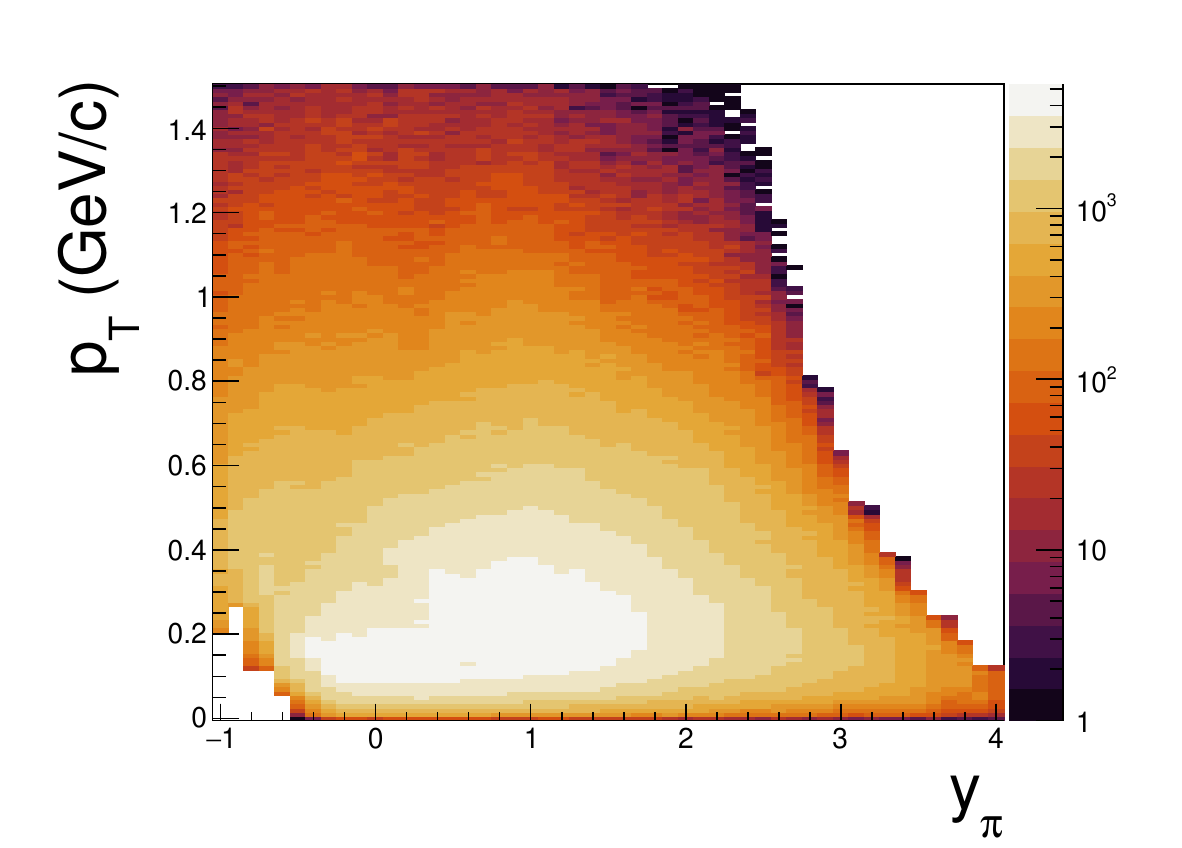}	\includegraphics[width=0.4\textwidth]{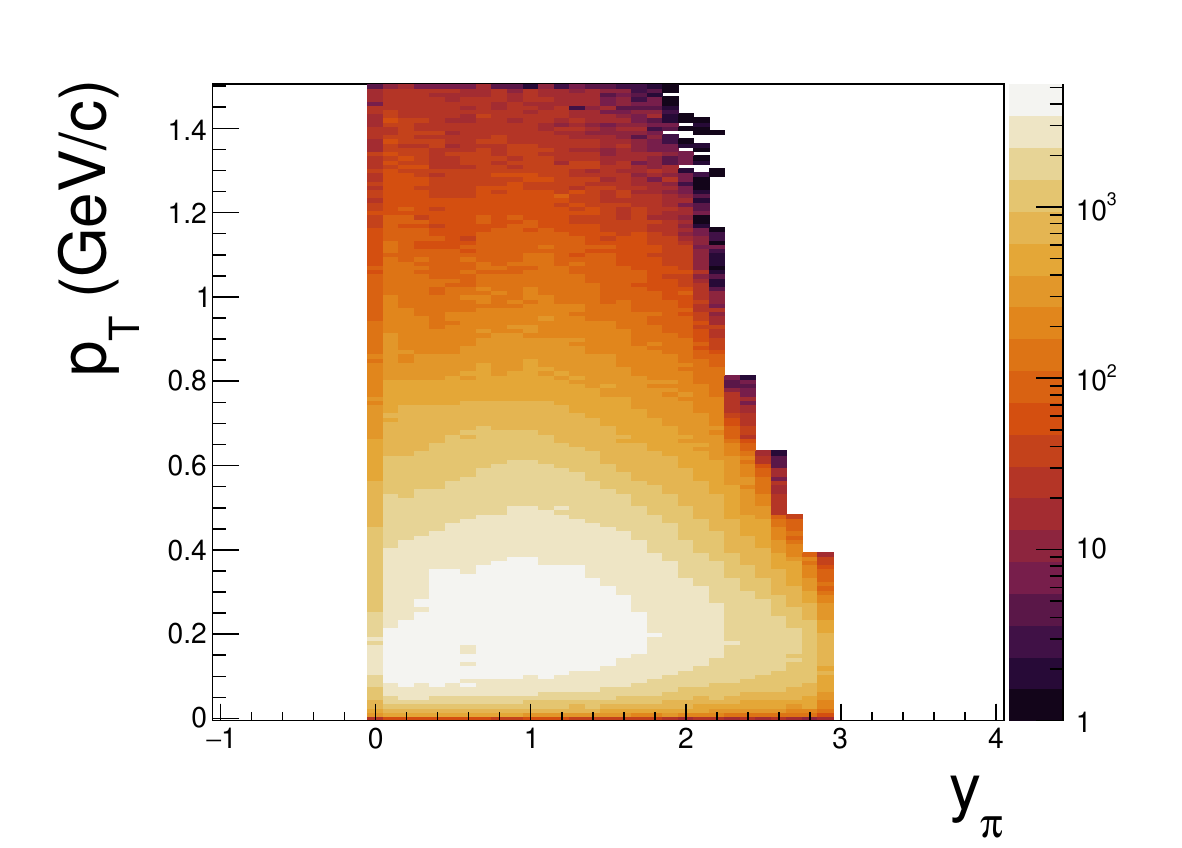}\\
			
			\includegraphics[width=0.4   \textwidth]{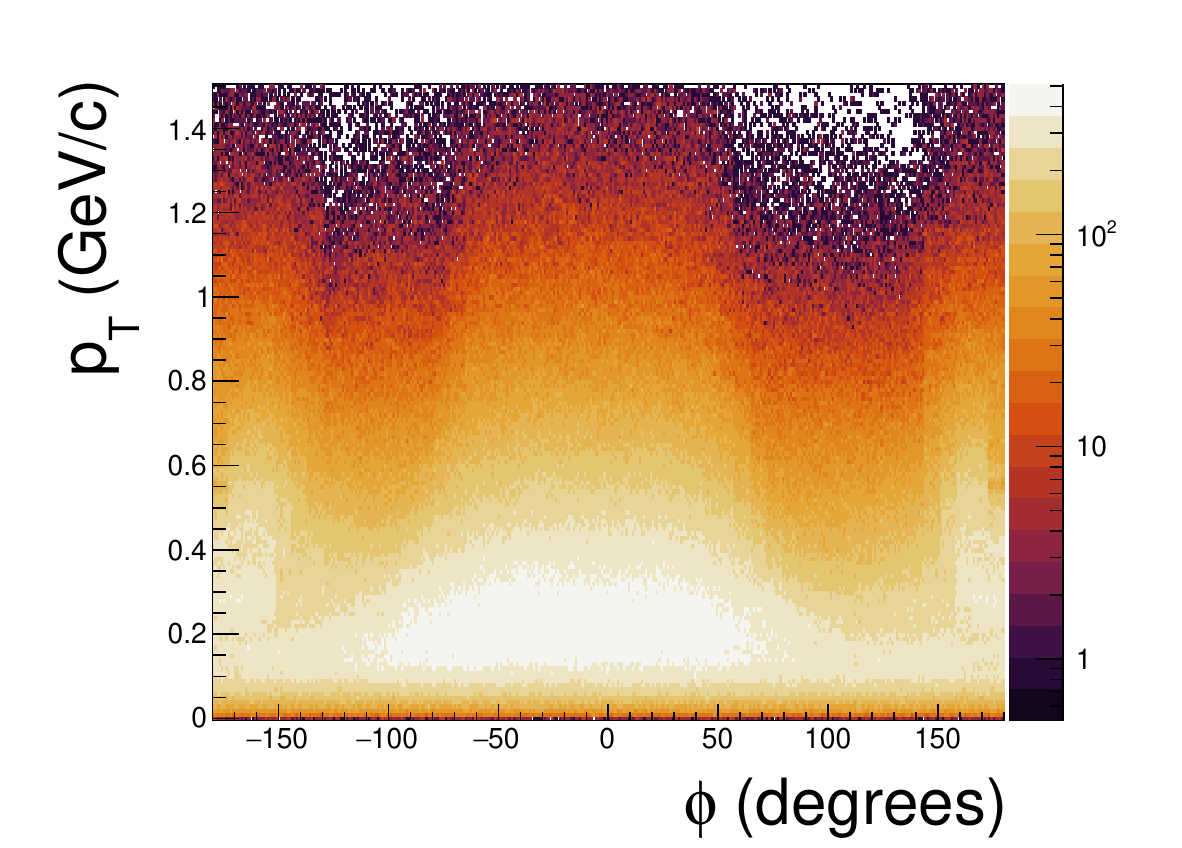}
			\includegraphics[width=0.4\textwidth]{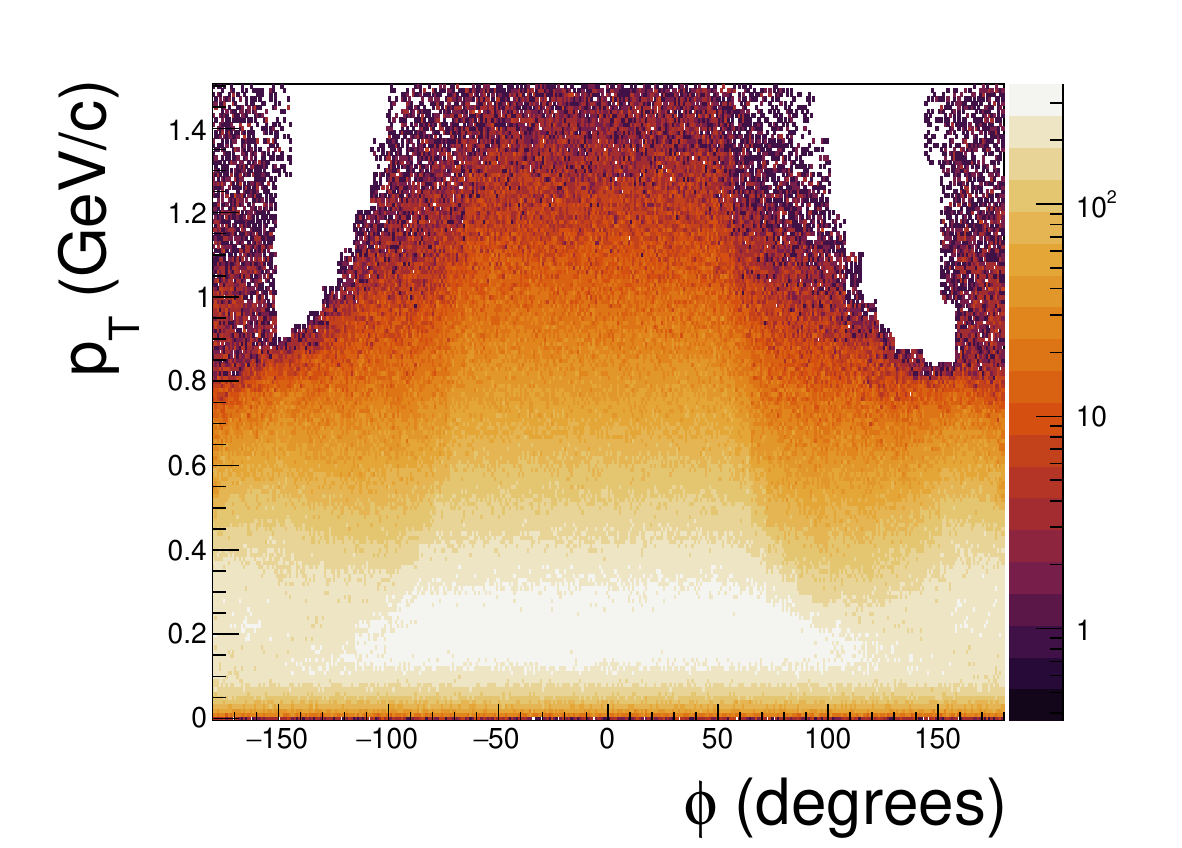}
			\caption[]{
				Analysis acceptances in $y_{\pi}-p_\text{T}$ and $\phi-p_\text{T}$ of charged hadrons in \pp interactions at 158~\GeVc used in Ref.~\cite{Aduszkiewicz:2015jna} (\textit{left}) and analysis acceptance in \pp interactions at 158~\GeVc common with nucleus-nucleus analysis~\cite{na61ArScAcc}~(\textit{right}).
			}
			\label{fig:acceptance}
		\end{figure}

Fluctuation results can not be corrected for limited analysis acceptance. 
In Ref.~\cite{Aduszkiewicz:2015jna}, fluctuations of multiplicity and transverse momentum up to second-order moments were analyzed in the entire high-quality phase-space region of the NA61/SHINE detector available for \pp interactions at a given beam momenta~\cite{na61ParticlePopulationMatrix}.
As the results of this analysis are planned to be compared to analysis results in nucleus-nucleus collisions, a different acceptance of the high-quality phase-space region of the NA61/SHINE detector in Ar+Sc interactions was used~\cite{na61ArScAcc}. Among others cut on upper $p_\text{T}<1.5$~\GeVc and rapidity of the track assuming pion mass $0<y_{\pi}<y_\text{beam}$, where $y_\text{beam}$ is the rapidity of the beam are included in Ar+Sc acceptance maps. The Figure~\ref{fig:acceptance} presents both acceptances in \pp interactions at 158~\GeVc.

Following Ref.~\cite{Savchuk:2019xfg} we calculate a fraction of accepted particles $x^{i}$ using the \EposLong~\cite{Werner:2008zza, Pierog:2009zt} model as:
\begin{equation}
    x^{i}=\frac{h_{acc}^{i}}{h^{i}},
\end{equation}
where $i$ stands for considered charge combination, $h_{acc}^{i}$ indicates the number of $i$-th particles in the analysis acceptance, and $h^{i}$ is the total number of $i$-th particles. 
The $x^{i}$ values for \pp interactions in \EposLong are given in Tab.~\ref{tab:acc}.

\begin{table}[]
    \centering
    \begin{tabular}{|c|c|c|c|c|}
    \hline
    $\sqrt{s_{NN}}$ (GeV)    & $h^{+}$ & $h^{-}$ & $h^{+}+h^{-}$  &  $h^{+}-h^{-}$  \\ 
    \hline
    6.3 &   0.24    &   0.26    &   0.25    &   0.23    \\
    7.7 &   0.25    &   0.28    &   0.26    &   0.23    \\
    8.8 &   0.26    &   0.30    &   0.27    &   0.23    \\
    12.3    &   0.30    &   0.36    &   0.32    &   0.24    \\
    17.3    &   0.35    &   0.43    &   0.38    &   0.26\\
    \hline
  \end{tabular}
    \caption{Fraction of the accepted charged hadrons and net-charge in \pp interactions within the analysis acceptance~\cite{na61ArScAcc} based on the \EposLong model. The fraction indicates analysis acceptance effects.}
    \label{tab:acc}
\end{table}


\subsection{Event and track selection}
Analyzed data consists only of events passing the trigger (Trigger in Tab.~\ref{tab:cuts}) condition. 
In the selected events, the trajectory of the beam particle was measured in at least one of BPD-1 or BPD-2 and in the BPD-3 detector (BPD in Tab.~\ref{tab:cuts}). 
To avoid bias from off-time events, an event is accepted only if it does not have the off-time beam particle within $\pm1$ $\mu$s around the trigger (beam) particle (WFA beam in Tab.~\ref{tab:cuts}). The main vertex \coordinate{z}-coordinate of the event has to be between $\pm 20$ cm around the center of the liquid hydrogen target~(fitted vertex \coordinate{z} position in Tab.~\ref{tab:cuts}). A small fraction of elastic events that pass the trigger condition (for beam momenta below 158 \GeVc) is removed by the removal of events with a single positive track with momentum close to beam momentum ($p\approx p_\text{beam}$ in Tab.~\ref{tab:cuts}). For details, see Ref.~\cite{NA61SHINE:2017fne}. A summary of the event selection (called \emph{standard cuts}) is given in the upper part of Table~\ref{tab:cuts}. 
The final number of events selected for the analysis is provided in Table~\ref{tab:events}.

\begin{table}[]
    \centering
    \begin{tabular}{|l|c|c|c|}
    \hline
    \cline{2-4}
    & standard cuts & loose cuts & tight cuts  \\ \hline
    Trigger & \multicolumn{3}{c|}{{ applied}}  \\ \cline{2-4}
    BPD & \multicolumn{3}{c|}{{ applied}} \\ \cline{2-4}
    WFA beam & $< \pm 1~\mu$s  & no cut & $< \pm 5 \mu$s \\ 
    fitted vertex \coordinate{z} position & $\pm 20$~cm & no cut & $\pm 10$~cm \\ \cline{2-4}
    $p \approx p_\text{beam}$ & \multicolumn{3}{c|}{applied} \\ \hline
    \hline
    total points & $\geq 30$ & no cut & $\geq 30$ \\ \cline{2-4}
    VTPC(GTPC) points & $\geq 15(5)$ & $\geq 10(5)$  & $\geq 15(5)$ \\ \cline{2-4}
    $|b_\text{x}|$ & $\leq 4$~cm & no cut & $\leq 4$~cm \\ \cline{2-4}
    $|b_\text{y}|$ & $\leq 2$~cm & no cut & $\leq 2$~cm \\ \cline{2-4}
    $e^{\pm}$ & \multicolumn{3}{c|}{applied} \\ \cline{2-4}
    acc map & \multicolumn{3}{c|}{applied} \\ \cline{2-4}
    \hline
  \end{tabular}
    \caption{Summary of event and track selection criteria used in the analysis. For details on cut definition, see text.}
    \label{tab:cuts}
\end{table}

\begin{table}[]
    \centering
    \begin{tabular}{|c|c|c|c|c|c|}
    \hline
    $\sqrt{s_{NN}}$ (GeV)    &  6.3 &   7.7 &   8.8 &   12.3    &   17.3 \\ \hline
    events &   218k &   928k    &   2.98M   &   1.67M   &   1.63M   \\
    \hline
  \end{tabular}
    \caption{Number of selected events after event selection.}
    \label{tab:events}
\end{table}

The above cuts allow the selection of good-quality inelastic events and the removal of remaining elastic scattering. The losses of inelastic interactions or bias of off-target interactions due to the event selection procedure were corrected for using simulation.

The selection of individual tracks was optimized to select hadrons produced in strong processes and electromagnetic decays. The selection ensured high reconstruction efficiency, proper identification of tracks, reduced contamination of tracks from secondary interactions,  weak decays, and off-time interactions. The following track selection criteria (called \emph{standard cuts})
were applied:
\begin{enumerate}[(i)]
\item the total number of reconstructed points on the track trajectory should be greater or equal 30 (total points in Tab.~\ref{tab:cuts}),
\item sum of the number of reconstructed points in VTPC-1 and VTPC-2 should be greater or equal to 15, or the number of reconstructed points in the GAP TPC should be greater or equal to 5 (VTPC(GTPC) points in Tab.~\ref{tab:cuts}),
\item distance between the track extrapolated to the interaction plane and the interaction point (track impact parameter)
should be smaller or equal 4~\cm in the horizontal (bending) plane and 2~\cm in the vertical (drift) plane ($|b_\text{x}|$ and $|b_\text{y}|$ in Tab.~\ref{tab:cuts}),
\item mean ionization energy loss measured for a given track does not indicate an electron or positron candidate as in Ref.~\cite{NA61SHINE:2013tiv}~($e^{\pm}$ in Tab.~\ref{tab:cuts}),
\item a track is measured in the high-efficiency region of the detector common with nucleus-nucleus analysis at a given beam momenta (acc map in Tab.~\ref{tab:cuts}).
\end{enumerate}

A summary of track selection criteria can be found in the lower part of Table~\ref{tab:cuts}. 







\subsection{Corrections}

Interactions with the target vessel and other materials in the target vicinity may contaminate the selected events. Also, inelastic events may be lost due to the limitations of the reconstruction procedure or detector. In the selected events, there may be contamination of hadrons coming from weak decays. In general, the distributions obtained using selected events and tracks may be affected by:
\begin{enumerate}[(i)]
	\item loss of inelastic events due to the online and offline event selection,
	\item contribution of elastic events, 
	\item contribution of off-target interactions, 
	\item loss of particles due to the detector and reconstruction inefficiency as well as due to track selection,
	\item contribution of particles from weak decays and secondary interactions (feed-down).
\end{enumerate}	
The unfolding procedure within RooUnfold library~\cite{unfolding} is used to correct the biases mentioned above. 
RooUnfold allows several methods of unfolding the distribution of interest, for example, bin-by-bin or iterative (\textit{Bayesian}) procedures. 
The iterative procedure was selected with seven iterations till the moment when the change of cumulant ratios with each step became much smaller than the statistical uncertainty for all considered distributions. 

Unfolding requires a description of the detector response, which was provided as a two-dimentional response matrix calculated using $N_\text{sim}$ and $N_\text{rec}$, where
\begin{enumerate}[(i)]
    \item $N_\text{sim}$ is the $N$ quantity obtained from simulated events and primary hadrons selected in the analysis acceptance,
    \item $N_\text{rec}$ is the $N$ quantity obtained from simulated events and primary hadrons after detector simulation and obtained in the same way as the reconstructed data.
\end{enumerate}
The response matrix is constructed in the FTFP-BERT~\cite{ALLISON2016186} model and the GEANT4~\cite{Agostinelli:2002hh,Allison:2006ve, Allison:2016lfl} detector response simulation. It is cross-checked with the response matrix obtained for the \EposLong~\cite{Werner:2008zza, Pierog:2009zt} model and GEANT3~\cite{Geant3} detector response simulation. The FTFP-BERT model is included in the GEANT4 framework and involves the following:
\begin{enumerate}[(i)]
   \item \textit{FTF} -- FRITIOF parton model,
    \item \textit{P} -- G4 Precompound model used for de-excitation of the remnant nucleus after
the initial high-energy interaction,    
   \item \textit{BERT} -- Bertini Cascade model.
\end{enumerate}
In the considered energy range the FTF model is invoked in the case of inelastic hadron processes. 
This specific solution was selected as it allows simulation of the passage of the beam particle through the target and detector setup. 
This way, one can address not only losses of inelastic events but also possible gains of elastic and off-target interactions (e.g., at the elements of the detector). 
The standard data-driven correction for off-target interactions usually applied to NA61/SHINE analysis~\cite{NA61SHINE:2017fne, Aduszkiewicz:2015jna} can not be used here due to limited statistics of removed-target data. The impact of non-target events after event selection in the FTFP-BERT remains below 6$\%$ in the studied reactions for the first moments of studied distributions close to what was estimated based on data-driven correction~\cite{NA61SHINE:2017fne, Aduszkiewicz:2015jna}. 

			 
Final distributions of the multiplicity of charged, positively, and negatively charged hadrons and net-charge refer to the unfolded ones. One-dimensional unfolding was applied in the case of positively and negatively charged hadron multiplicity distributions. In the case of the multiplicity distribution of all charged hadrons and net-charge distribution, two options were considered:
\begin{enumerate}[(i)]
    \item correction of the one-dimensional distribution of $h^{+}\pm h^{-}$, or,
    \item correction of the two-dimensional distribution of $h^{+}$ and $h^{-}$.
\end{enumerate}   
In general, 2D unfolding considers correlations between positively and negatively charged hadrons. As it involves 2D distributions, it requires larger statistics than 1D. In this analysis, both approaches were tested with models (one model was treated as the data, and the other was used for the unfolding) and with data (comparison of 1D vs. 2D results). Both provided the same results within statistical uncertainties. Thus, for all charged and net-charge, one-dimensional unfolding was used. Multiplicity and net-charge distributions provide natural binning as the number of hadrons is quantized 
and the bin size is fixed to unity.


\subsection{Statistical uncertainty}

Intensive quantities are constructed as ratios of cumulants of a distribution. To account for correlations between cumulants, statistical uncertainty was obtained using the bootstrap method~\cite{efron, moore}. The method requires constructing artificial data sets (S-sets) of the same size as the data. We have constructed 100 bootstrap samples. 
All analysis steps were performed for each bootstrap sample. Thus, S-sets contain bootstraped data and a response matrix. The final uncertainty is then calculated as the standard deviation of the distribution of a given quantity obtained from all S-sets. 

\subsection{Systematic uncertainty}

Systematic uncertainties originate from the imperfectness of the detector response/reconstruction procedure 
and uncertainties induced by the description of physical processes implemented in the models.
The total systematic uncertainties were obtained by selecting the most significant effect, either variation of event and track selection or selection of the model. The largest bias is taken symmetrically.

The detector-related effects were addressed by varying event and track selections -- see Table~\ref{tab:cuts} for the definition of event and track selection variation. 
So-called \emph{tight} and \emph{loose} selections refer to extreme scenarios of loose and tight data selection. The loose set of cuts was defined following Ref.~\cite{Aduszkiewicz:2015jna}. The tight selection definition differs from the standard selection only in event selection. 
The tight track selection was kept the same as in the case of standard selection, not to add the change of acceptance bias. Both data sets (loose and tight) were corrected the same way as the standard data. 

\begin{figure*}[h]
			\centering
			\includegraphics[width=0.8\textwidth]{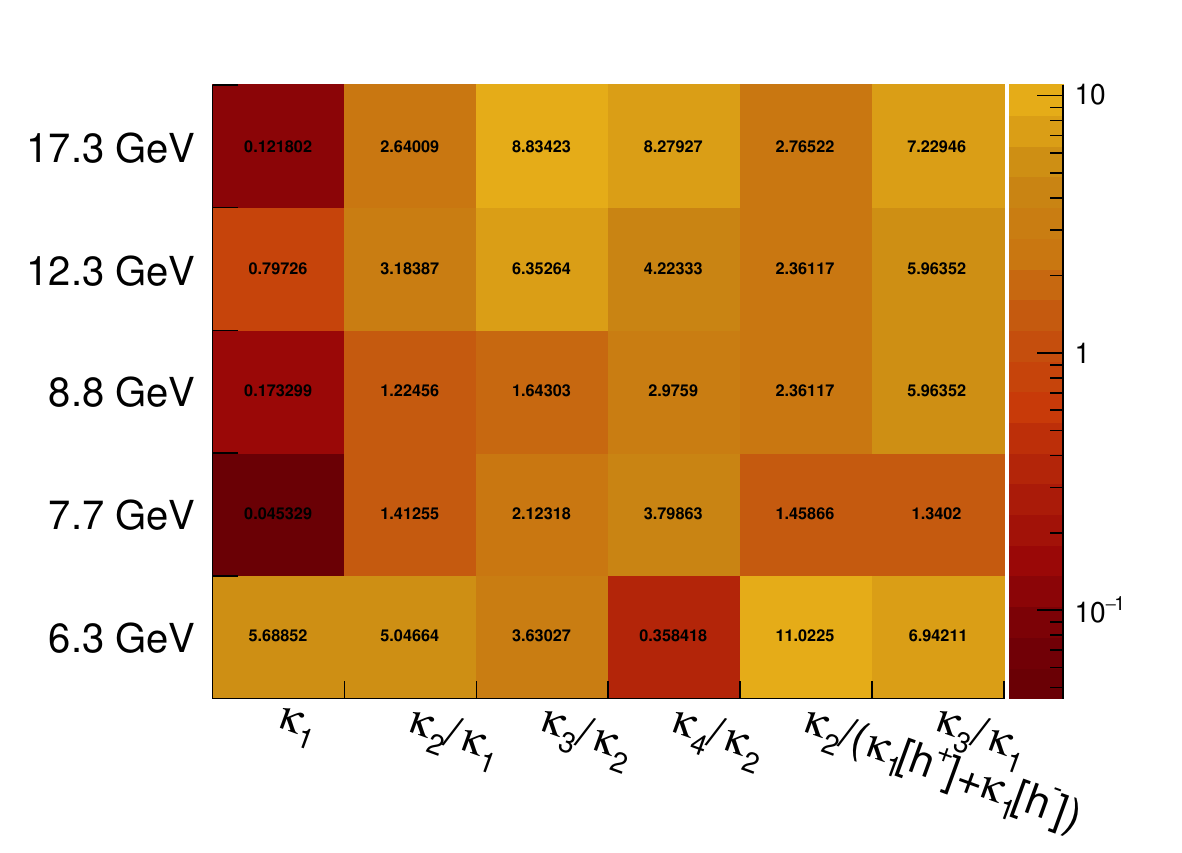}	
			\caption[]{
			Difference (in $\%$) of unfolded and true values of quantities of the net-charge distribution where a sample from the \EposLong model was treated as data (with statistics close to the experimental data) and the response matrix was built from FTFP-BERT.
			}
			\label{fig:1DMCTest}
		\end{figure*}

The model-related uncertainties originate from the imperfectness of the model used to unfold distributions.  
The uncertainties were estimated using simulations performed within the FTFP-BERT and \EposLong models. As a check, the simulated \EposLong data were corrected using corrections based on the FTFP-BERT model and compared to the unbiased \EposLong results. 
In this check, the unfolding always improved agreement between the obtained results and the true ones. 
An example of the test of unfolding in the case of net-charge distribution in \pp interactions in the \EposLong model is shown in Fig.~\ref{fig:1DMCTest}. 

 A dominant source of uncertainty becomes apparent from the comparison of different event and track selection
criteria. On average, this uncertainty is 3.2 times larger than that emerging from the comparison of unfolding procedures using different
models. In less than $20\%$ cases, the dominating uncertainty emerges from the comparison of unfolding with \EposLong and
FTFP-BERT models -- this takes place usually in the case of single cumulants and rarely in the case of ratios. For the ratio of $\kappa_4/\kappa_2$ the dominant effect is nearly always the comparison of the event and track selection (the exception is the $h^{-}$ at the lowest energy where the effects are comparable).

\section{Results}
\label{sec:Results}
This section presents results on multiplicity and net-charge fluctuations of charged primary hadrons in inelastic \pp interactions at $\sqrt{s_{NN}}=6.3, 7.7, 8.8, 12.3$, and 17.3 \GeV. In the first subsection, final corrected distributions of the considered charge combinations are presented along with model predictions and raw measured distributions, including detector effects.
The second and third subsections
present results on intensive quantities, which allow for a direct comparison with nucleus-nucleus collisions, and on factorial cumulants, which allow studying correlations.

\subsection{Multiplicity and net-charge distributions}
 \begin{figure*}
			\centering
  			\includegraphics[width=\textwidth]{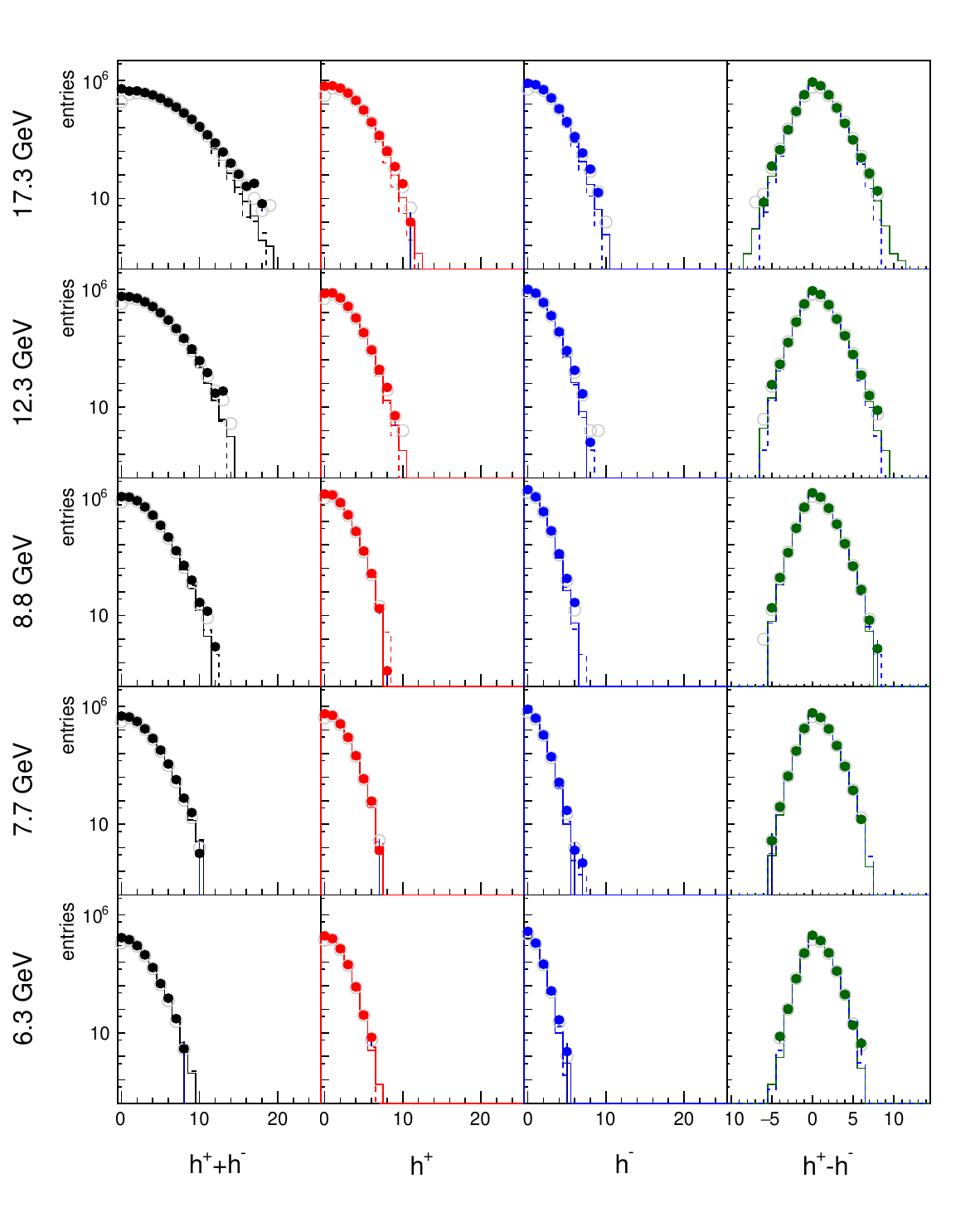}	
			\caption[]{
			Multiplicity distributions of $h^{+}+h^{-}$, $h^{+}$, $h^{-}$, and $h^{+}-h^{-}$ in \pp interactions at $\sqrt{s_{NN}}=6.3$, 7.7, 8.8, 12.3, and 17.3 \GeV in the phase-space region as defined in Ref.~\cite{na61ArScAcc}. Open circles indicate raw, uncorrected data, whereas full circles show corrected distributions. Dashed and solid lines represent FTFP-BERT and \EposLong model predictions, respectively.    
			}
			\label{fig:distr}
		\end{figure*}
Figure~\ref{fig:distr} shows corrected distributions of $h^{+}+h^{-}$, $h^{+}$, $h^{-}$, and $h^{+}-h^{-}$ (full circles) along with the raw measured distributions (open circles). The experimental results are
compared to FTFP-BERT and \EposLong models predictions (dashed and solid lines). 
		
\subsection{Intensive quantities}
    \begin{figure*}
			\centering
			\includegraphics[width=\textwidth]{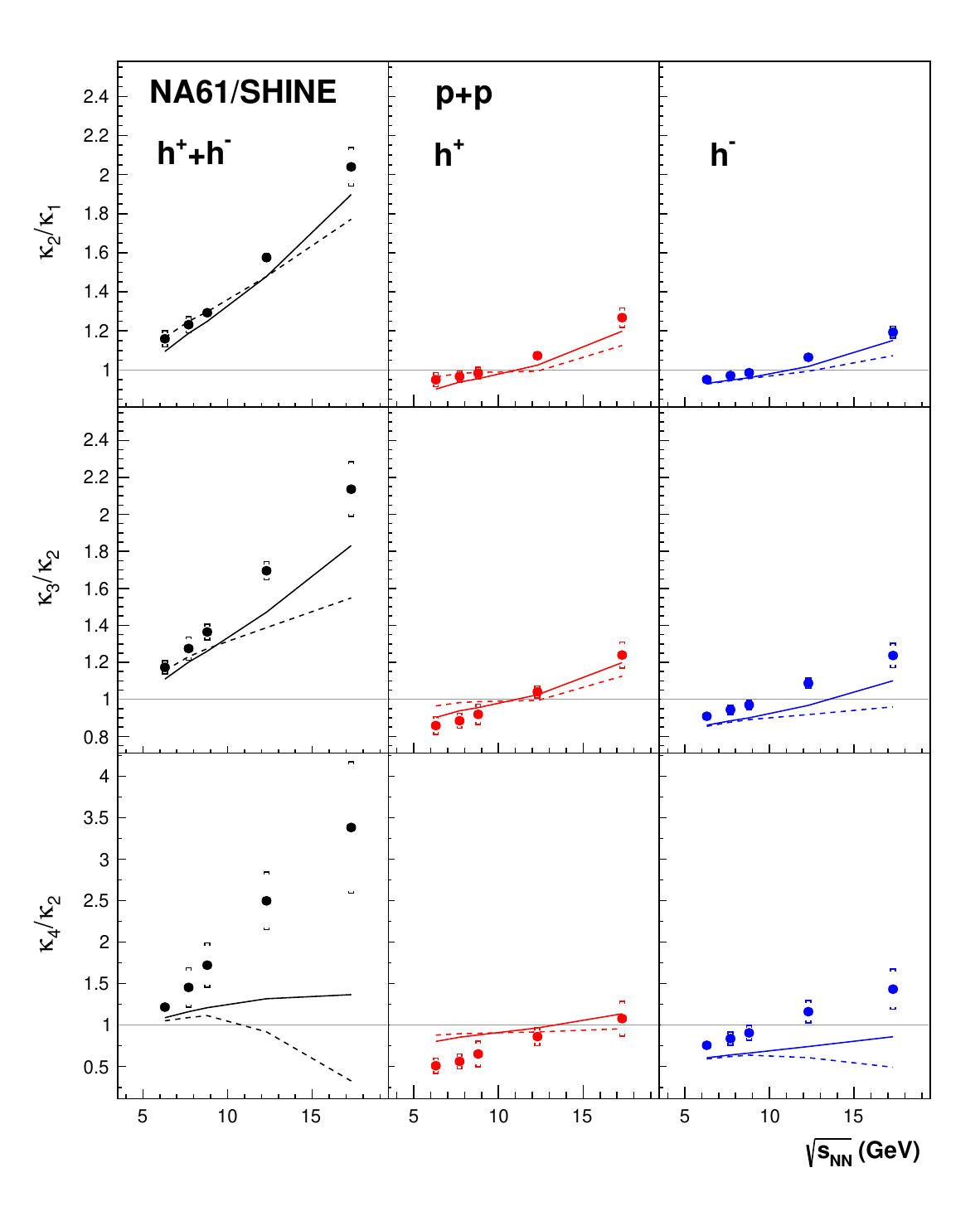}	
			\caption[]{Energy dependence of intensive quantities of multiplicity distributions of $h^{+}+h^{-}$, $h^{+}$, and $h^{-}$ in \pp interactions at $\sqrt{s_{NN}}=6.3$, 7.7, 8.8, 12.3, and 17.3 \GeV in the phase-space region as defined in Ref.~\cite{na61ArScAcc}. The statistical uncertainty is indicated with a color bar (often smaller than the marker), and systematic uncertainty is indicated with a square bracket. Results are compared with \EposLong (solid line) and FTFP-BERT (dashed line) predictions.}
			\label{fig:results_mult}
		\end{figure*}
		
		Results on the energy dependence of intensive quantities are presented in Fig.~\ref{fig:results_mult}. All quantities increase with the interaction energy. The increase is the strongest for the sum of charges.
  All quantities obtained for summed charged hadrons remain above unity at all considered collision energies. This implies that fluctuations are enhanced with respect to Poisson.
  The same energy dependence can be observed in the case of positively 
  and negatively charged hadrons, but the increase is much weaker with the signal crossing unity around $\sqrt{s_{NN}}\approx 10$ GeV except $\kappa_{4}/\kappa_{2}[h^{+}]$ where stays close to one for higher energies. 
		The rise with interaction energy stays in agreement with the KNO-G scaling~\cite{Koba:1972ng,Golokhvastov:2001ei,Golokhvastov:2001pt} observed in full kinematic acceptance~\cite{Heiselberg:2000fk}. 
		However, another reason for the increasing strength of the signal may be the increase of the analysis acceptance with interaction energy (see Table~\ref{tab:acc}).
		
		The results are compared with \EposLong and FTFP-BERT models. The \EposLong model is a multiple scattering approach based on partons and Pomerons (parton ladders)~\cite{Pierog:2009zt}. The FTFP-BERT was already utilized for performing the unfolding procedure as described in Sec.~\ref{sec:analysis}. It is a combination of string hadronic models used by the GEANT4 framework and described in details in Ref.~\cite{ALLISON2016186}. Both models reproduce the experimental $\kappa_{2}/\kappa_{1}$ ratio but tend to underestimate $\kappa_{3}/\kappa_{2}$ and qualitatively disagree with $\kappa_{4}/\kappa_{2}$ for $h^{+}+h^{-}$ and $h^{-}$. It is unclear why, in the case of FTFP-BERT predictions, there is a maximum close to $\sqrt{s_{NN}}\approx 9$ GeV in the case of all and negatively charged hadrons.
		\begin{figure*}
			\centering
	    	\includegraphics[width=\textwidth]{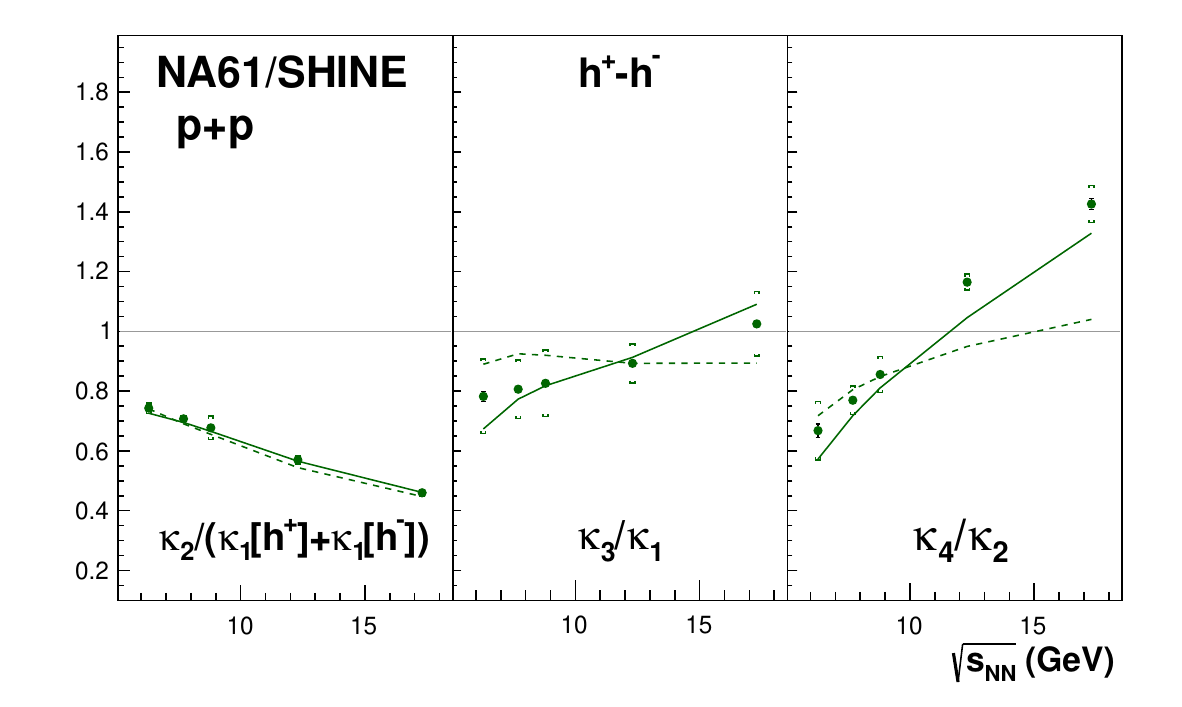}
			\caption[]{
			Energy dependence of intensive quantities of net-charge distribution in \pp interactions at $\sqrt{s_{NN}}=6.3$, 7.7, 8.8, 12.3, and 17.3 \GeV in the phase-space region as defined in Ref.~\cite{na61ArScAcc}. The statistical uncertainty is indicated with a color bar (often smaller than the marker), and systematic uncertainty is indicated with a square bracket. Results are compared with \EposLong (solid line) and FTFP-BERT (dashed line) predictions.
			}
			\label{fig:results_net}
		\end{figure*}
		
		Figure~\ref{fig:results_net} shows the energy dependence of net-charge fluctuations compared with model predictions. Second-order cumulant ratios of $h^{+}-h^{-}$ distribution decreases with collision energy, whereas $\kappa_{3}[h^{+}-h^{-}]/\kappa_{1}[h^{+}-h^{-}]$ and $\kappa_{4}[h^{+}-h^{-}]/\kappa_{2}[h^{+}-h^{-}]$ increase with collision energy. The measured signal for the majority of collected energies remains below unity. Both model predictions reproduce the observed $\kappa_{2}[h^{+}-h^{-}]/(\kappa_{1}[h^{+}]+\kappa_{1}[h^{-}])$ and magnitude of $\kappa_{3}[h^{+}-h^{-}]/\kappa_{1}[h^{+}-h^{-}]$. Only $\kappa_{4}[h^{+}-h^{-}]/\kappa_{2}[h^{+}-h^{-}]$ at the two top energies is higher than unity. The \EposLong model reproduces this rise with interaction energy whereas FTFP-BERT underestimates its strength. 
  
\subsection{Factorial cumulants}
        Factorial cumulants are quantities that allow extracting the correlation function of a given order (and without lower-order terms) from the measured distribution~\cite{Bzdak:2016sxg, Kitazawa:2017ljq}. 
		\begin{figure*}
			\centering
			\includegraphics[width=\textwidth]{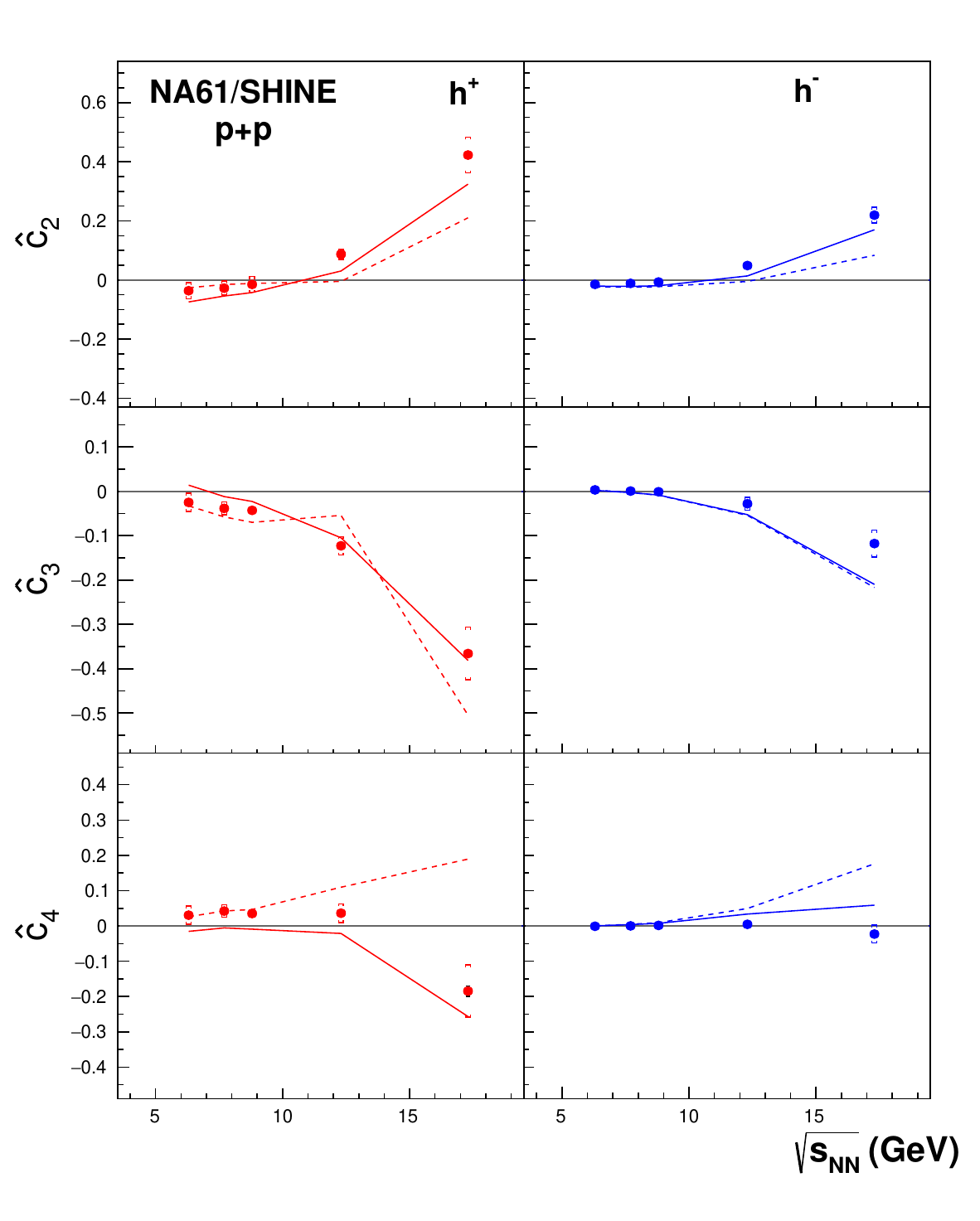}
			\caption[]{
			Energy dependence of factorial cumulants of multiplicity distributions of $h^{+}$ and $h^{-}$ in \pp interactions at $\sqrt{s_{NN}}=6.3$, 7.7, 8.8, 12.3, and 17.3 \GeV in the phase-space region as defined in Ref.~\cite{na61ArScAcc}. The statistical uncertainty is indicated with a color bar (often smaller than the marker), and systematic uncertainty is indicated with a square bracket. Results are compared with \EposLong (solid line) and FTFP-BERT (dashed line) predictions.
			}
			\label{fig:results_C}
		\end{figure*}	
		The energy dependence of factorial cumulants measured in \pp interactions is presented in Fig.~\ref{fig:results_C}. 
  The $\hat{C}_{2}$ signal of $h^{+}$ and $h^{-}$ distributions slowly increases from values close to zero to positive values at top energy. 
  In contrast, $\hat{C}_{3}$ of $h^{+}$ and $h^{-}$ and $\hat{C}_{4}$ of $h^{+}$ decrease with interaction energy. For $h^{+}$ the magnitude of the decrease is larger for $\hat{C}_{3}$ than for $\hat{C}_{4}$.
        Both models generally describe the trends and magnitudes of the measured factorial cumulants, except the FTFP-BERT model, which does not describe the energy dependence of $\hat{C}_{4}$.

\section{Comparison with other systems}

Quantitative comparison between systems is possible only if they were performed in a similar acceptance~\cite{Vovchenko:2020tsr}. We leave such a comparison for future analysis of nucleus-nucleus collisions from the system-size scan of \NASixtyOne.

Nevertheless, qualitative comparisons between different experiments may provide useful information. The \NASixtyOne results on multiplicity fluctuations (studied with $\kappa_{2}/\kappa_{1}$ ratio) were already reported and discussed in Ref.~\cite{Aduszkiewicz:2015jna}. Higher-order cumulant ratios of multiplicity distributions reported here are the first results provided 
in \pp interactions for the considered energy range. 

\begin{figure}
    \centering
    \includegraphics[width=0.42\textwidth]{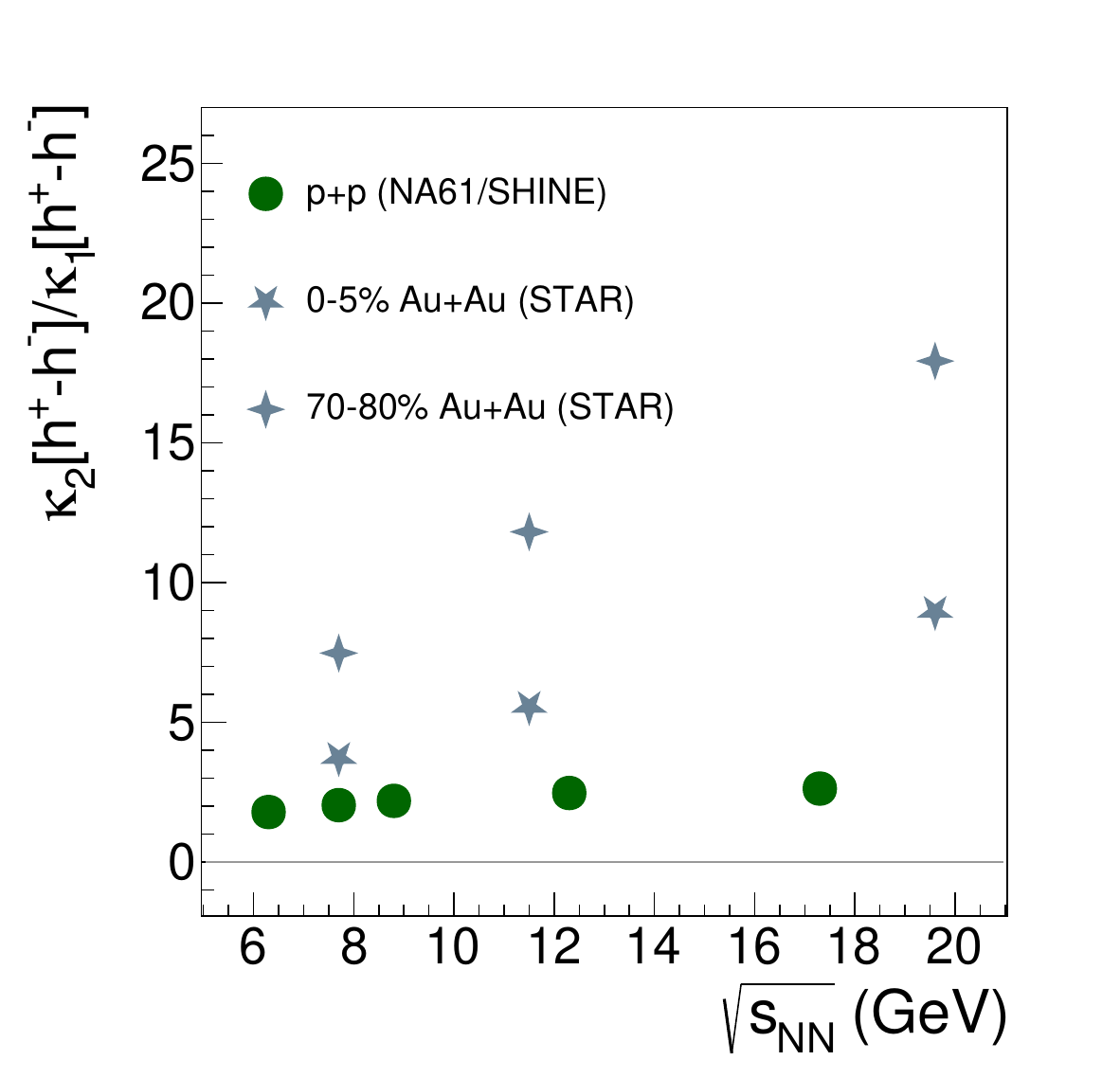}\includegraphics[width=0.42\textwidth]{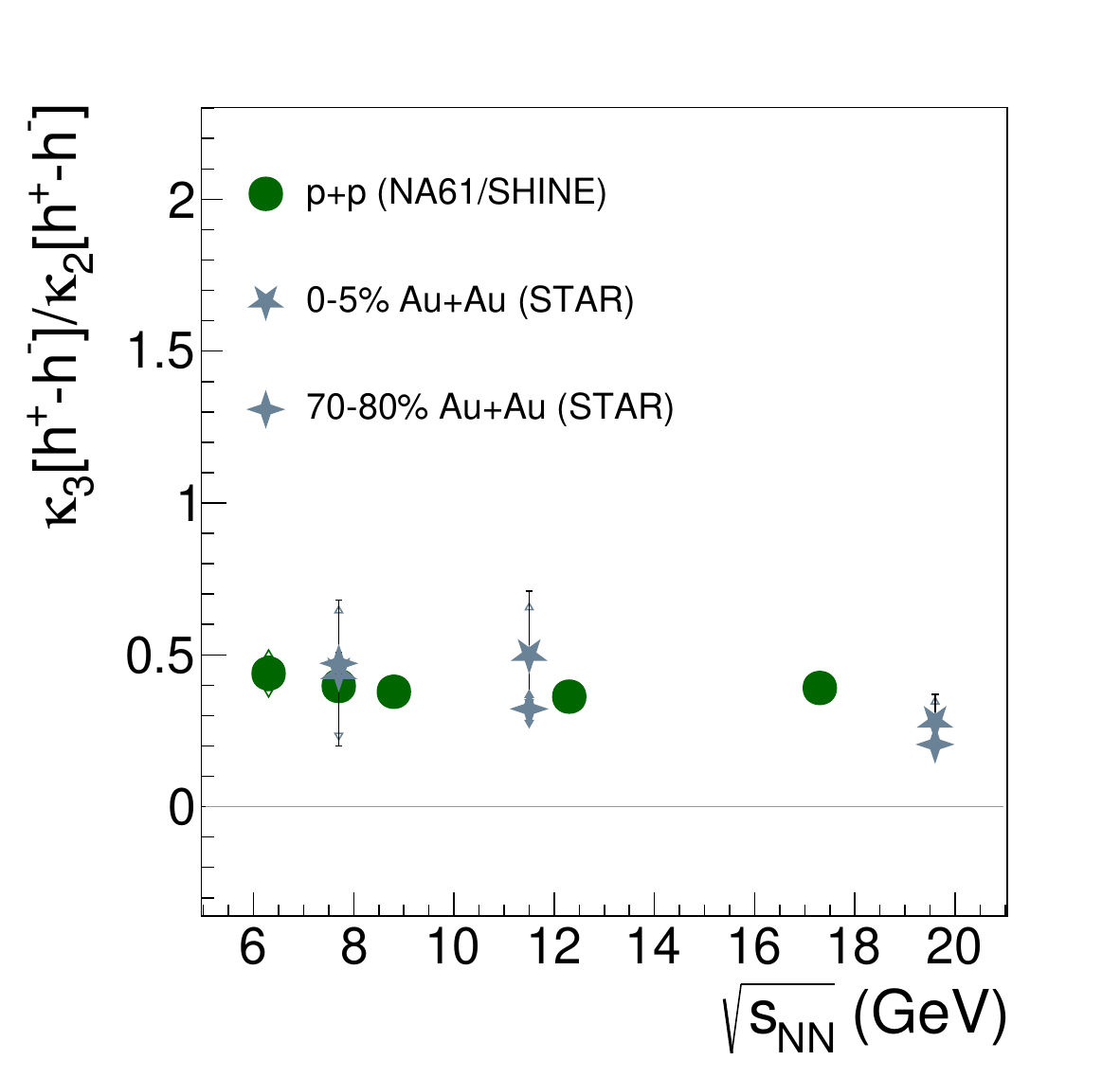}\\
    \includegraphics[width=0.42\textwidth]{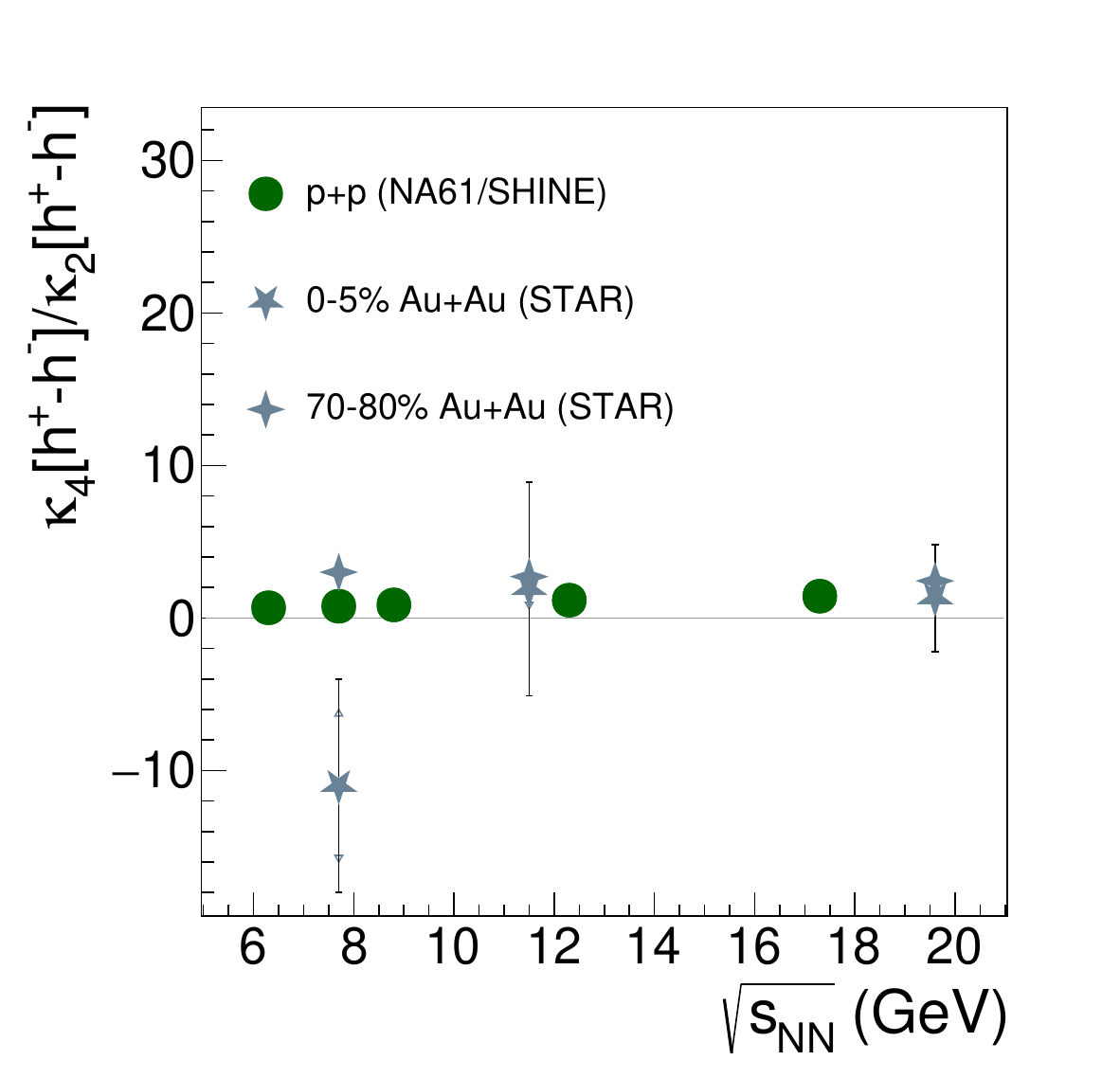}
    
    \caption{ Net-charge cumulant ratios measured in \pp interactions by \NASixtyOne and in central and peripheral Au+Au interactions by STAR~\cite{STAR:2014egu}. }
    \label{fig:STAR_net}
\end{figure}

Results on the net-charge distribution were compared to 
results from the NA49~\cite{NA49:2004fqq} and STAR~\cite{STAR:2014egu,Pandav:2020uzx} experiments. 
\begin{figure}
    \centering
    \begin{minipage}{0.45\textwidth}
    \includegraphics[width=\textwidth]{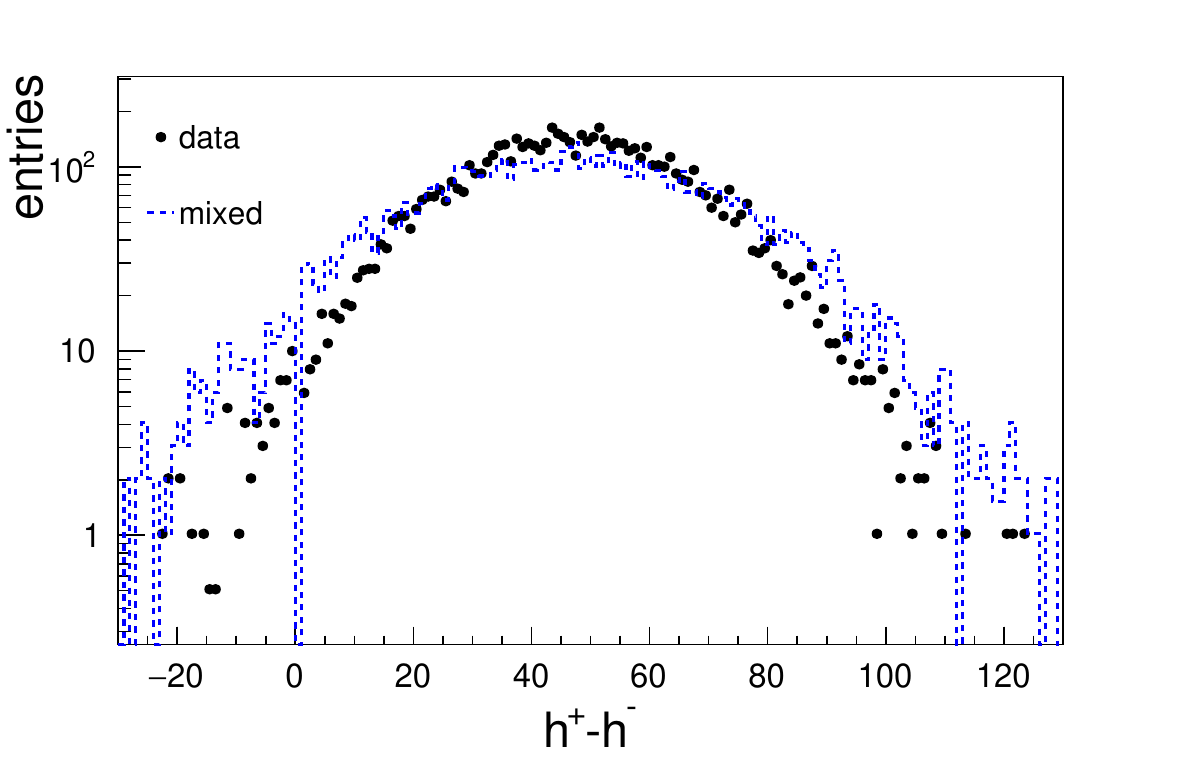}
    \end{minipage}
    \begin{minipage}{0.45\textwidth}
     \includegraphics[width=\textwidth]{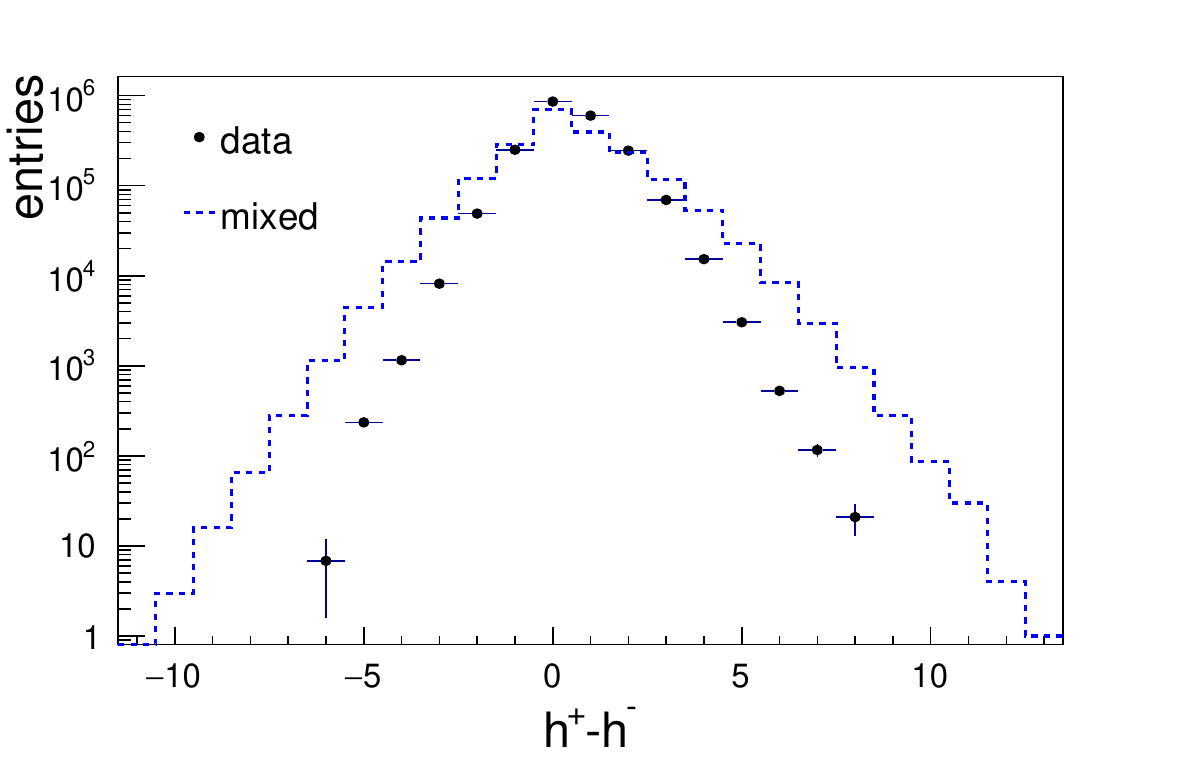}
    \end{minipage}
    
    \caption{Net-charge distribution in the 10$\%$ most central Pb+Pb interactions by NA49~(reproduced with the data from Ref.~\cite{NA49:2004fqq}) (\textit{left}) and in \pp interactions by \NASixtyOne~(\textit{right}) at $\sqrt{s_{NN}}=$ 17.3 \GeV. Both distributions are compared with mixed events (blue line) as defined in Ref.~\cite{NA49:2004fqq}.}
    \label{fig:na49_net}
\end{figure}
The left panel of Fig.~\ref{fig:na49_net} presents net-charge distribution measured by NA49~\cite{NA49:2004fqq} in the 10$\%$ most central Pb+Pb interactions at beam momentum 158\AGeVc. The NA49 analysis acceptance is comparable with the one used in this analysis. The NA49 data distribution (points) is compared with mixed events (blue line), which are constructed by randomly selecting particles from different events according to the multiplicity distribution measured for the data. The set of mixed events was prepared in the same way for \pp interactions (see the right panel in Fig.~\ref{fig:na49_net}). In both systems, mixed events distribution is wider than the data thus both distributions seem to be dominated by conservation laws. The net-charge distribution (around 7k central Pb+Pb interactions) provided by NA49 allows for estimating values of cumulant ratios for this reaction. Using formulas for statistical uncertainty estimation from Ref.~\cite{Luo:2011tp}, one may try to provide NA49 results with approximate statistical uncertainties. Thus, cumulant ratios in central Pb+Pb interactions at 158$A$ \GeVc are  $\kappa_{2}[h^{+}-h^{-}]/\kappa_{1}[h^{+}-h^{-}]=8.16\pm 0.17$~(stat), 
$\kappa_{3}[h^{+}-h^{-}]/\kappa_{2}[h^{+}-h^{-}]=0.9\pm 1.6$ (stat), and $\kappa_{4}[h^{+}-h^{-}]/\kappa_{2}[h^{+}-h^{-}]=21\pm32$ (stat), approximately. 
The obtained ratios, within large uncertainties, are not too far from results in the 5$\%$ most central Au+Au collisions at $\sqrt{s_{NN}}=$
19.6 GeV~\cite{STAR:2014egu}; see Fig.~\ref{fig:STAR_net}. 
The similarity may indicate that the acceptance difference is not too large for a qualitative comparison. On the other hand, one should remember that it may be coincidental to some degree. Moreover, volume fluctuations or resonance contributions may be different in both systems.  

The comparison of \pp collisions with Au+Au interactions~\cite{STAR:2014egu} is presented in Fig.~\ref{fig:STAR_net}. It should be underlined that quantities $\kappa_{2}[h^{+}-h^{-}]/\kappa_{1}[h^{+}-h^{-}]$ and $\kappa_{3}[h^{+}-h^{-}]/\kappa_{2}[h^{+}-h^{-}]$ do not keep 0 and 1 as their reference values (see Sec.~\ref{sec:IQ}). 
Instead, for Skellam distribution $\kappa_{2}[h^{+}-h^{-}]/\kappa_{1}[h^{+}-h^{-}]$ should increase and $\kappa_{3}[h^{+}-h^{-}]/\kappa_{2}[h^{+}-h^{-}]$ should decrease with increasing multiplicity. 
In the case of scaled variance, \pp interactions are well below central and peripheral Au+Au interactions.
The difference may be caused by volume fluctuations, which are unavoidable with wide centrality bins. Additionally, it can also result from acceptance differences in the phase space of the analysis (see discussion in Ref.~\cite{Anticic:2013htn}). This includes changes in the acceptance of \pp analysis with collision energy.
In GCE, volume fluctuations are modulated by the mean number of particles produced in a fixed $V$. 
Thus, the scaled variance should increase with increasing collision energy as more particles are produced at higher energies, explaining the observed energy dependence for Au+Au reactions. Such fluctuations
are absent in the case of \pp collisions, explaining a weaker increase with interaction energy. 

The observed signal of $\kappa_{3}[h^{+}-h^{-}]/\kappa_{2}[h^{+}-h^{-}]$ and $\kappa_{4}[h^{+}-h^{-}]/\kappa_{2}[h^{+}-h^{-}]$ in Au+Au is close to \pp interactions. Although volume fluctuations dependence of higher-order cumulant ratios remains, it is more elaborate (see Ref.~\cite{Begun:2016sop}) and seems not to dominate the signal. 

\section{Summary}
\label{sec:summary}

The experimental results on event-by-event
fluctuations of multiplicities of all, positively, and negatively charged hadrons as well as net-electric charge (so-called net-charge) produced in inelastic proton-proton interactions
at  $\!\sqrt{s_{\mathrm{NN}}}$ = 6.3, 7.7, 8.8, 12.3, and 17.3~\GeV are presented.
The results were corrected for experimental biases with the unfolding technique. The corrected results were compared with \EposLong and FTFP-BERT predictions. In general, both models qualitatively describe the measurements. The \EposLong predictions are closer to the data. The KNO-G scaling and an increase in the analysis acceptance with collision energy probably caused the rise of cumulant ratios with the collision energy for multiplicity distributions ($h^{+}+h^{-}$, $h^{+}$, and $h^{-}$). The qualitative disagreement of FTFP-BERT for higher collision energies in $\kappa_{4}/\kappa_{2}$ can be seen for $h^{+}+h^{-}$ and $h^{-}$.

The most significant deviation of the measured signal to model predictions appears mostly for the sum of charges, indicating possible problems of models with describing correlations between charges, like resonances and conservation laws.

A qualitative comparison with Au+Au interactions was performed. The measured signals are not far from each other but one should keep in mind differences in considered phase space as well as volume fluctuations in Au+Au. Future comparisons are expected with precise NA61/SHINE results on nucleus-nucleus collisions.
\appendix
\section{Appendix A}

The numerical values of corrected distributions will be provided using the HEP Data~\cite{Maguire:2017ypu, HEPData}. The numerical values of measured quantities are shown in Tabs.~\ref{tab:cumRatios},~\ref{tab:table_factcum},~\ref{tab:net2}. The first uncertainty is statistical and the second is systematic. It should be underlined that $\kappa_{1}$ is not equivalent to the values measured by particle yields~\cite{NA61SHINE:2017fne, NA61SHINE:2013tiv} due to different event and track selections as well as different correction procedures. 

\begin{table}[]
    \centering
    \tiny{
    \begin{tabular}{c|c|c|c|c|c}
         $\sqrt{s_{NN}}$ (GeV) & quantity &   $h^{+}+h^{-}$ &   $h^{+}$ &   $h^{-}$ & $h^{+}-h^{-}$\\
         \hline
         \hline
    6.3	&	$\kappa_{1}$	& $	1.0222	\pm	0.0027	\pm	0.071	$ & $	0.7242	\pm	0.0020	\pm	0.038	$ & $	0.3027	\pm	0.0012	\pm	0.025	$ & $	0.4270	\pm	0.0022	\pm	0.012	$ \\
7.7	&	$\kappa_{1}$	& $	1.2381	\pm	0.0015	\pm	0.055	$ & $	0.8365	\pm	0.0011	\pm	0.031	$ & $	0.40468	\pm	0.00070	\pm	0.018	$ & $	0.4325	\pm	0.0012	\pm	0.011	$ \\
8.8	&	$\kappa_{1}$	& $	1.3913	\pm	0.0010	\pm	0.16	$ & $	0.91188	\pm	0.00063	\pm	0.096	$ & $	0.48169	\pm	0.00043	\pm	0.057	$ & $	0.43230	\pm	0.00068	\pm	0.033	$ \\
12.3	&	$\kappa_{1}$	& $	1.9373	\pm	0.0017	\pm	0.12	$ & $	1.1922	\pm	0.0010	\pm	0.066	$ & $	0.75133	\pm	0.00077	\pm	0.048	$ & $	0.4493	\pm	0.0010	\pm	0.014	$ \\
17.3	&	$\kappa_{1}$	& $	2.6839	\pm	0.0025	\pm	0.11	$ & $	1.5781	\pm	0.0014	\pm	0.055	$ & $	1.1348	\pm	0.0011	\pm	0.039	$ & $	0.4758	\pm	0.0011	\pm	0.0061	$ \\
\hline
6.3	&	$\kappa_{2}$	& $	1.1854	\pm	0.0047	\pm	0.042	$ & $	0.6879	\pm	0.0023	\pm	0.014	$ & $	0.2879	\pm	0.0014	\pm	0.023	$ & $	0.7614	\pm	0.0032	\pm	0.030	$ \\
7.7	&	$\kappa_{2}$	& $	1.5250	\pm	0.0029	\pm	0.033	$ & $	0.8088	\pm	0.0013	\pm	0.016	$ & $	0.39328	\pm	0.00090	\pm	0.013	$ & $	0.8777	\pm	0.0017	\pm	0.024	$ \\
8.8	&	$\kappa_{2}$	& $	1.7987	\pm	0.0019	\pm	0.19	$ & $	0.89692	\pm	0.00092	\pm	0.070	$ & $	0.47492	\pm	0.00059	\pm	0.053	$ & $	0.9436	\pm	0.0010	\pm	0.056	$ \\
12.3	&	$\kappa_{2}$	& $	3.0518	\pm	0.0041	\pm	0.15	$ & $	1.2795	\pm	0.0019	\pm	0.049	$ & $	0.8002	\pm	0.0013	\pm	0.046	$ & $	1.1072	\pm	0.0017	\pm	0.037	$ \\
17.3	&	$\kappa_{2}$	& $	5.4738	\pm	0.0067	\pm	0.15	$ & $	2.0008	\pm	0.0027	\pm	0.065	$ & $	1.3541	\pm	0.0018	\pm	0.025	$ & $	1.2476	\pm	0.0020	\pm	0.041	$ \\
\hline
6.3	&	$\kappa_{3}$	& $	1.391	\pm	0.015	\pm	0.034	$ & $	0.5903	\pm	0.0056	\pm	0.024	$ & $	0.2616	\pm	0.0028	\pm	0.019	$ & $	0.3340	\pm	0.0060	\pm	0.044	$ \\
7.7	&	$\kappa_{3}$	& $	1.944	\pm	0.010	\pm	0.071	$ & $	0.7150	\pm	0.0037	\pm	0.026	$ & $	0.3712	\pm	0.0020	\pm	0.010	$ & $	0.3488	\pm	0.0039	\pm	0.033	$ \\
8.8	&	$\kappa_{3}$	& $	2.4540	\pm	0.0071	\pm	0.19	$ & $	0.8241	\pm	0.0024	\pm	0.021	$ & $	0.4605	\pm	0.0014	\pm	0.040	$ & $	0.3571	\pm	0.0024	\pm	0.018	$ \\
12.3	&	$\kappa_{3}$	& $	5.173	\pm	0.021	\pm	0.13	$ & $	1.3313	\pm	0.0055	\pm	0.049	$ & $	0.8699	\pm	0.0035	\pm	0.028	$ & $	0.4012	\pm	0.0047	\pm	0.027	$ \\
17.3	&	$\kappa_{3}$	& $	11.692	\pm	0.042	\pm	0.85	$ & $	2.480	\pm	0.010	\pm	0.15	$ & $	1.6751	\pm	0.0064	\pm	0.072	$ & $	0.4877	\pm	0.0064	\pm	0.056	$ \\
\hline
6.3	&	$\kappa_{4}$	& $	1.442	\pm	0.057	\pm	0.041 $ & $	0.351	\pm	0.015	\pm	0.056	$ & $	0.2179	\pm	0.0075	\pm	0.015	$ & $	0.534	\pm	0.021	\pm	0.052	$ \\
7.7	&	$\kappa_{4}$	& $	2.217	\pm	0.043	\pm	0.33	$ & $	0.455	\pm	0.010	\pm	0.067	$ & $	0.3292	\pm	0.0050	\pm	0.019	$ & $	0.675	\pm	0.011	\pm	0.022	$ \\
8.8	&	$\kappa_{4}$	& $	3.098	\pm	0.012	\pm	0.10	$ & $	0.5851	\pm	0.0087	\pm	0.082	$ & $	0.4304	\pm	0.0041	\pm	0.018	$ & $	0.8074	\pm	0.0078	\pm	 0.022 $ \\
12.3	&	$\kappa_{4}$	& $	7.62	\pm	0.12	\pm	0.66	$ & $	1.103	\pm	0.019	\pm	0.094	$ & $	0.930	\pm	0.019	\pm	0.053	$ & $	1.289	\pm	0.015	\pm	0.054	$ \\
17.3	&	$\kappa_{4}$	& $	18.51	\pm	0.31	\pm	4.4	$ & $	2.157	\pm	0.041	\pm	0.43	$ & $	1.941	\pm	0.028	\pm	0.31	$ & $	1.779	\pm	0.025	\pm	0.13	$ \\
\hline
6.3	&	$\kappa_{2}/\kappa_{1}$	& $	1.1596	\pm	0.0040	\pm	0.041	$ & $	0.9499	\pm	0.0028	\pm	0.036	$ & $	0.9511	\pm	0.0030	\pm	0.0051	$ & $	1.7831	\pm	0.0094	\pm	0.028	$ \\
7.7	&	$\kappa_{2}/\kappa_{1}$	& $	1.2317	\pm	0.0019	\pm	0.042	$ & $	0.9669	\pm	0.0014	\pm	0.028	$ & $	0.9718	\pm	0.0014	\pm	0.012	$ & $	2.0295	\pm	0.0054	\pm	0.011	$ \\
8.8	&	$\kappa_{2}/\kappa_{1}$	& $	1.2929	\pm	0.0011	\pm	0.014	$ & $	0.98359	\pm	0.00092	\pm	0.031	$ & $	0.98596	\pm	0.00088	\pm	0.010	$ & $	2.1828	\pm	0.0036	\pm	0.039	$ \\
12.3	&	$\kappa_{2}/\kappa_{1}$	& $	1.5753	\pm	0.0021	\pm	0.021	$ & $	1.0732	\pm	0.0014	\pm	0.019	$ & $	1.0650	\pm	0.0013	\pm	0.0070	$ & $	2.4641	\pm	0.0049	\pm	0.067	$ \\
17.3	&	$\kappa_{2}/\kappa_{1}$	& $	2.0395	\pm	0.0027	\pm	0.098	$ & $	1.2678	\pm	0.0017	\pm	0.049	$ & $	1.1933	\pm	0.0013	\pm	0.032	$ & $	2.6221	\pm	0.0062	\pm	0.057	$ \\
\hline
6.3	&	$\kappa_{3}/\kappa_{2}$	& $	1.1731	\pm	0.0093	\pm	0.037	$ & $	0.8582	\pm	0.0064	\pm	0.048	$ & $	0.9086	\pm	0.0068	\pm	0.0081	$ & $	0.4387	\pm	0.0075	\pm	0.077	$ \\
7.7	&	$\kappa_{3}/\kappa_{2}$	& $	1.2748	\pm	0.0048	\pm	0.063	$ & $	0.8841	\pm	0.0035	\pm	0.041	$ & $	0.9438	\pm	0.0034	\pm	0.026	$ & $	0.3973	\pm	0.0043	\pm	0.050	$ \\
8.8	&	$\kappa_{3}/\kappa_{2}$	& $	1.3643	\pm	0.0029	\pm	0.042	$ & $	0.9189	\pm	0.0020	\pm	0.054	$ & $	0.9696	\pm	0.0020	\pm	0.025	$ & $	0.3785	\pm	0.0026	\pm	0.044	$ \\
12.3	&	$\kappa_{3}/\kappa_{2}$	& $	1.6952	\pm	0.0051	\pm	0.049	$ & $	1.0405	\pm	0.0032	\pm	0.032	$ & $	1.0871	\pm	0.0031	\pm	0.028	$ & $	0.3624	\pm	0.0042	\pm	0.028	$ \\
17.3	&	$\kappa_{3}/\kappa_{2}$	& $	2.1360	\pm	0.0061	\pm	0.14	$ & $	1.2396	\pm	0.0039	\pm	0.071	$ & $	1.2371	\pm	0.0036	\pm	0.065	$ & $	0.3909	\pm	0.0051	\pm	0.033	$ \\

\hline
6.3	&	$\kappa_{4}/\kappa_{2}$	& $	1.217	\pm	0.045	\pm	0.023	$ & $	0.510	\pm	0.024	\pm	0.092	$ & $	0.757	\pm	0.024	\pm	0.013	$ & $	0.668	\pm	0.022	\pm	0.098	$ \\
7.7	&	$\kappa_{4}/\kappa_{2}$	& $	1.454	\pm	0.028	\pm	0.24	$ & $	0.562	\pm	0.013	\pm	0.089	$ & $	0.837	\pm	0.013	\pm	0.079	$ & $	0.769	\pm	0.013	\pm	0.048	$ \\
8.8	&	$\kappa_{4}/\kappa_{2}$	& $	1.722	\pm	0.019	\pm	0.27	$ & $	0.6523	\pm	0.0087	\pm	0.15	$ & $	0.9064	\pm	0.0087	\pm	0.090	$ & $	0.8557	\pm	0.0082	\pm	0.060	$ \\
12.3	&	$\kappa_{4}/\kappa_{2}$	& $	2.498	\pm	0.040	\pm	0.35	$ & $	0.862	\pm	0.015	\pm	0.11	$ & $	1.162	\pm	0.015	\pm	0.14	$ & $	1.165	\pm	0.014	\pm	0.027	$ \\
17.3	&	$\kappa_{4}/\kappa_{2}$	& $	3.382	\pm	0.058	\pm	0.79	$ & $	1.078	\pm	0.022	\pm	0.21	$ & $	1.433	\pm	0.020	\pm	0.24	$ & $	1.426	\pm	0.018	\pm	0.062	$ \\
\hline
\hline
    \end{tabular}
    }

    \caption{Numerical values of cumulants and cumulant ratios of $h^{+}+h^{-}$, $h^{+}$, $h^{-}$, and $h^{+}-h^{-}$.}
    \label{tab:cumRatios}
\end{table}

\begin{table}[]
    \centering
    \tiny{
\begin{tabular}{c|c|c|c}
         $\sqrt{s_{NN}}$ (GeV) & quantity &     $h^{+}$ &   $h^{-}$ \\
         \hline
    \hline
6.3	&	$\hat{C_{2}}$	&  $	-0.0363	\pm	0.0021	\pm	0.027	$ & $	-0.01481	\pm	0.00090	\pm	0.0026	$  \\
7.7	&	$\hat{C_{2}}$	&  $	-0.0277	\pm	0.0012	\pm	0.024	$ & $	-0.01140	\pm	0.00056	\pm	0.0051	$ \\
8.8	&	$\hat{C_{2}}$	&  $	-0.01496	\pm	0.00084	\pm	0.027	$ & $	-0.00676	\pm	0.00042	\pm	0.0049	$ \\
12.3	&	$\hat{C_{2}}$	& $	0.0873	\pm	0.0017	\pm	0.018	$ & $	0.0489	\pm	0.0010	\pm	0.0017	$ \\
17.3	&	$\hat{C_{2}}$	& $	0.4226	\pm	0.0026	\pm	0.060	$ & $	0.2194	\pm	0.0015	\pm	0.027	$ \\
\hline
6.3	&	$\hat{C_{3}}$	& $	-0.0249	\pm	0.0028	\pm	0.020	$ & $	0.00331	\pm	0.00075	\pm	0.0011	$ \\
7.7	&	$\hat{C_{3}}$	& $	-0.0384	\pm	0.0018	\pm	0.014	$ & $	0.00069	\pm	0.00072	\pm	0.00024	$ \\
8.8	&	$\hat{C_{3}}$	& $	-0.0429	\pm	0.0013	\pm	0.0031	$ & $	-0.00089	\pm	0.00051	\pm	0.0026	$ \\
12.3	&	$\hat{C_{3}}$	& $	-0.1228	\pm	0.0028	\pm	0.021	$ & $	-0.0280	\pm	0.0017	\pm	0.014	$ \\
17.3	&	$\hat{C_{3}}$	& $	-0.3658	\pm	0.0056	\pm	0.059	$ & $	-0.1177	\pm	0.0033	\pm	0.030	$ \\
\hline
6.3	&	$\hat{C_{4}}$	& $	0.0304	\pm	0.0048	\pm		0.026 $ & $	-0.00100	\pm	0.00076	\pm	0.0013	$ \\
7.7	&	$\hat{C_{4}}$	& $	0.0421 \pm	0.0027	\pm	0.018	$ & $	0.00012	\pm	0.00082	\pm	0.00043	$ \\
8.8	&	$\hat{C_{4}}$	& $	0.0351	\pm	0.0021	\pm	0.0090	$ & $	0.0014	\pm	0.0010	\pm	0.0014	$ \\
12.3	&	$\hat{C_{4}}$	& $	0.0361	\pm	0.0056	\pm	0.027	$ & $	0.0045	\pm	0.0026	\pm	0.0028	$ \\
17.3	&	$\hat{C_{4}}$	& $	-0.185	\pm	0.015	\pm	0.074	$ & $	-0.0231	\pm	0.0084	\pm	0.025	$ \\
\hline
\hline
\end{tabular}
    }

    \caption{Numerical values of factorial cumulants of $h^{+}$ and $h^{-}$.}
    \label{tab:table_factcum}
\end{table}

\begin{table}[]
    \centering
    \begin{tabular}{c|c|c}
    $\sqrt{s_{NN}}$ (GeV)& quantity & $h^{+}-h^{-}$\\
         \hline
             \hline
6.3	&	$\kappa_{2}[h^{+}-h^{-}]/(\kappa_{1}[h^{+}]+\kappa_{1}[h^{-}])$	& $	0.7431	\pm	0.0032	\pm	0.017	$ \\
7.7	&	$\kappa_{2}[h^{+}-h^{-}]/(\kappa_{1}[h^{+}]+\kappa_{1}[h^{-}])$	& $	0.7072	\pm	0.0017	\pm	0.0088	$ \\
8.8	&	$\kappa_{2}[h^{+}-h^{-}]/(\kappa_{1}[h^{+}]+\kappa_{1}[h^{-}])$	& $	0.6771	\pm	0.0010	\pm	0.039	$ \\
12.3	&	$\kappa_{2}[h^{+}-h^{-}]/(\kappa_{1}[h^{+}]+\kappa_{1}[h^{-}])$	& $	0.5697	\pm	0.0017	\pm	0.014	$ \\
17.3	&	$\kappa_{2}[h^{+}-h^{-}]/(\kappa_{1}[h^{+}]+\kappa_{1}[h^{-}])$	& $	0.4599	\pm	0.0020	\pm	0.010	$ \\
         \hline
6.3	&	$\kappa_{3}[h^{+}-h^{-}]/\kappa_{1}[h^{+}-h^{-}]$	& $	0.782	\pm	0.017	\pm		0.13$ \\
7.7	&	$\kappa_{3}[h^{+}-h^{-}]/\kappa_{1}[h^{+}-h^{-}]$	& $	0.8064	\pm	0.0085	\pm	0.098	$ \\
8.8	&	$\kappa_{3}[h^{+}-h^{-}]/\kappa_{1}[h^{+}-h^{-}]$	& $	0.8264	\pm	0.0058	\pm		0.11$ \\
12.3	&	$\kappa_{3}[h^{+}-h^{-}]/\kappa_{1}[h^{+}-h^{-}]$	& $	0.893	\pm	0.011	\pm	0.065	$ \\
17.3	&	$\kappa_{3}[h^{+}-h^{-}]/\kappa_{1}[h^{+}-h^{-}]$	& $	1.025 \pm	0.012	\pm	0.11	$ \\
\hline
\hline
    \end{tabular}
    \caption{Numerical values of cumulant ratio combinations which are intensive for net-charge.}
    \label{tab:net2}
\end{table}

\clearpage
\section*{Acknowledgements}
We would like to thank the CERN EP, BE, HSE and EN Departments for the
strong support of NA61/SHINE.

This work was supported by
the Hungarian Scientific Research Fund (grant NKFIH 138136\slash137812\slash138152 and TKP2021-NKTA-64),
the Polish Ministry of Science and Higher Education
(DIR\slash WK\slash\-2016\slash 2017\slash\-10-1, WUT ID-UB), the National Science Centre Poland (grants
2014\slash 14\slash E\slash ST2\slash 00018, 
2016\slash 21\slash D\slash ST2\slash 01983, 
2017\slash 25\slash N\slash ST2\slash 02575, 
2018\slash 29\slash N\slash ST2\slash 02595, 
2018\slash 30\slash A\slash ST2\slash 00226, 
2018\slash 31\slash G\slash ST2\slash 03910, 
2020\slash 39\slash O\slash ST2\slash 00277), 
the Norwegian Financial Mechanism 2014--2021 (grant 2019\slash 34\slash H\slash ST2\slash 00585),
the Polish Minister of Education and Science (contract No. 2021\slash WK\slash 10),
the European Union's Horizon 2020 research and innovation programme under grant agreement No. 871072,
the Ministry of Education, Culture, Sports,
Science and Tech\-no\-lo\-gy, Japan, Grant-in-Aid for Sci\-en\-ti\-fic
Research (grants 18071005, 19034011, 19740162, 20740160 and 20039012,22H04943),
the German Research Foundation DFG (grants GA\,1480\slash8-1 and project 426579465),
the Bulgarian Ministry of Education and Science within the National
Roadmap for Research Infrastructures 2020--2027, contract No. D01-374/18.12.2020,
Serbian Ministry of Science, Technological Development and Innovation (grant
OI171002), Swiss Nationalfonds Foundation (grant 200020\-117913/1),
ETH Research Grant TH-01\,07-3, National Science Foundation grant
PHY-2013228 and the Fermi National Accelerator Laboratory (Fermilab),
a U.S. Department of Energy, Office of Science, HEP User Facility
managed by Fermi Research Alliance, LLC (FRA), acting under Contract
No. DE-AC02-07CH11359 and the IN2P3-CNRS (France).\\

The data used in this paper were collected before February 2022.

\bibliographystyle{include/na61Utphys}


\bibliography{include/na61References}

\newpage
{\Large The \NASixtyOne Collaboration}
\bigskip
\begin{sloppypar}

\noindent
{H.\;Adhikary~\href{https://orcid.org/0000-0002-5746-1268}{\includegraphics[height=1.7ex]{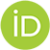}}\textsuperscript{\,11}},
{P.\;Adrich~\href{https://orcid.org/0000-0002-7019-5451}{\includegraphics[height=1.7ex]{orcid-logo.png}}\textsuperscript{\,13}},
{K.K.\;Allison~\href{https://orcid.org/0000-0002-3494-9383}{\includegraphics[height=1.7ex]{orcid-logo.png}}\textsuperscript{\,24}},
{N.\;Amin~\href{https://orcid.org/0009-0004-7572-3817}{\includegraphics[height=1.7ex]{orcid-logo.png}}\textsuperscript{\,4}},
{E.V.\;Andronov~\href{https://orcid.org/0000-0003-0437-9292}{\includegraphics[height=1.7ex]{orcid-logo.png}}\textsuperscript{\,20}},
{I.-C.\;Arsene~\href{https://orcid.org/0000-0003-2316-9565}{\includegraphics[height=1.7ex]{orcid-logo.png}}\textsuperscript{\,10}},
{M.\;Bajda~\href{https://orcid.org/0009-0005-8859-1099}{\includegraphics[height=1.7ex]{orcid-logo.png}}\textsuperscript{\,14}},
{Y.\;Balkova~\href{https://orcid.org/0000-0002-6957-573X}{\includegraphics[height=1.7ex]{orcid-logo.png}}\textsuperscript{\,16}},
{D.\;Battaglia~\href{https://orcid.org/0000-0002-5283-0992}{\includegraphics[height=1.7ex]{orcid-logo.png}}\textsuperscript{\,23}},
{A.\;Bazgir~\href{https://orcid.org/0000-0003-0358-0576}{\includegraphics[height=1.7ex]{orcid-logo.png}}\textsuperscript{\,11}},
{S.\;Bhosale~\href{https://orcid.org/0000-0001-5709-4747}{\includegraphics[height=1.7ex]{orcid-logo.png}}\textsuperscript{\,12}},
{M.\;Bielewicz~\href{https://orcid.org/0000-0001-8267-4874}{\includegraphics[height=1.7ex]{orcid-logo.png}}\textsuperscript{\,13}},
{A.\;Blondel~\href{https://orcid.org/0000-0002-1597-8859}{\includegraphics[height=1.7ex]{orcid-logo.png}}\textsuperscript{\,3}},
{M.\;Bogomilov~\href{https://orcid.org/0000-0001-7738-2041}{\includegraphics[height=1.7ex]{orcid-logo.png}}\textsuperscript{\,2}},
{Y.\;Bondar~\href{https://orcid.org/0000-0003-2773-9668}{\includegraphics[height=1.7ex]{orcid-logo.png}}\textsuperscript{\,11}},
{A.\;Borucka\textsuperscript{\,19}},
{A.\;Brandin\textsuperscript{\,20}},
{W.\;Bryli\'nski~\href{https://orcid.org/0000-0002-3457-6601}{\includegraphics[height=1.7ex]{orcid-logo.png}}\textsuperscript{\,19}},
{J.\;Brzychczyk~\href{https://orcid.org/0000-0001-5320-6748}{\includegraphics[height=1.7ex]{orcid-logo.png}}\textsuperscript{\,14}},
{M.\;Buryakov~\href{https://orcid.org/0009-0008-2394-4967}{\includegraphics[height=1.7ex]{orcid-logo.png}}\textsuperscript{\,20}},
{A.F.\;Camino\textsuperscript{\,26}},
{M.\;\'Cirkovi\'c~\href{https://orcid.org/0000-0002-4420-9688}{\includegraphics[height=1.7ex]{orcid-logo.png}}\textsuperscript{\,21}},
{M.\;Csan\'ad~\href{https://orcid.org/0000-0002-3154-6925}{\includegraphics[height=1.7ex]{orcid-logo.png}}\textsuperscript{\,6}},
{J.\;Cybowska~\href{https://orcid.org/0000-0003-2568-3664}{\includegraphics[height=1.7ex]{orcid-logo.png}}\textsuperscript{\,19}},
{T.\;Czopowicz~\href{https://orcid.org/0000-0003-1908-2977}{\includegraphics[height=1.7ex]{orcid-logo.png}}\textsuperscript{\,11}},
{C.\;Dalmazzone~\href{https://orcid.org/0000-0001-6945-5845}{\includegraphics[height=1.7ex]{orcid-logo.png}}\textsuperscript{\,3}},
{N.\;Davis~\href{https://orcid.org/0000-0003-3047-6854}{\includegraphics[height=1.7ex]{orcid-logo.png}}\textsuperscript{\,12}},
{A.\;Dmitriev~\href{https://orcid.org/0000-0001-7853-0173}{\includegraphics[height=1.7ex]{orcid-logo.png}}\textsuperscript{\,20}},
{P.~von\;Doetinchem~\href{https://orcid.org/0000-0002-7801-3376}{\includegraphics[height=1.7ex]{orcid-logo.png}}\textsuperscript{\,25}},
{W.\;Dominik~\href{https://orcid.org/0000-0001-7444-9239}{\includegraphics[height=1.7ex]{orcid-logo.png}}\textsuperscript{\,17}},
{J.\;Dumarchez~\href{https://orcid.org/0000-0002-9243-4425}{\includegraphics[height=1.7ex]{orcid-logo.png}}\textsuperscript{\,3}},
{R.\;Engel~\href{https://orcid.org/0000-0003-2924-8889}{\includegraphics[height=1.7ex]{orcid-logo.png}}\textsuperscript{\,4}},
{G.A.\;Feofilov~\href{https://orcid.org/0000-0003-3700-8623}{\includegraphics[height=1.7ex]{orcid-logo.png}}\textsuperscript{\,20}},
{L.\;Fields~\href{https://orcid.org/0000-0001-8281-3686}{\includegraphics[height=1.7ex]{orcid-logo.png}}\textsuperscript{\,23}},
{Z.\;Fodor~\href{https://orcid.org/0000-0003-2519-5687}{\includegraphics[height=1.7ex]{orcid-logo.png}}\textsuperscript{\,5,18}},
{M.\;Friend~\href{https://orcid.org/0000-0003-4660-4670}{\includegraphics[height=1.7ex]{orcid-logo.png}}\textsuperscript{\,7}},
{M.\;Ga\'zdzicki~\href{https://orcid.org/0000-0002-6114-8223}{\includegraphics[height=1.7ex]{orcid-logo.png}}\textsuperscript{\,11}},
{O.\;Golosov~\href{https://orcid.org/0000-0001-6562-2925}{\includegraphics[height=1.7ex]{orcid-logo.png}}\textsuperscript{\,20}},
{V.\;Golovatyuk~\href{https://orcid.org/0009-0006-5201-0990}{\includegraphics[height=1.7ex]{orcid-logo.png}}\textsuperscript{\,20}},
{M.\;Golubeva~\href{https://orcid.org/0009-0003-4756-2449}{\includegraphics[height=1.7ex]{orcid-logo.png}}\textsuperscript{\,20}},
{K.\;Grebieszkow~\href{https://orcid.org/0000-0002-6754-9554}{\includegraphics[height=1.7ex]{orcid-logo.png}}\textsuperscript{\,19}},
{F.\;Guber~\href{https://orcid.org/0000-0001-8790-3218}{\includegraphics[height=1.7ex]{orcid-logo.png}}\textsuperscript{\,20}},
{S.N.\;Igolkin\textsuperscript{\,20}},
{S.\;Ilieva~\href{https://orcid.org/0000-0001-9204-2563}{\includegraphics[height=1.7ex]{orcid-logo.png}}\textsuperscript{\,2}},
{A.\;Ivashkin~\href{https://orcid.org/0000-0003-4595-5866}{\includegraphics[height=1.7ex]{orcid-logo.png}}\textsuperscript{\,20}},
{A.\;Izvestnyy~\href{https://orcid.org/0009-0009-1305-7309}{\includegraphics[height=1.7ex]{orcid-logo.png}}\textsuperscript{\,20}},
{N.\;Kargin\textsuperscript{\,20}},
{N.\;Karpushkin~\href{https://orcid.org/0000-0001-5513-9331}{\includegraphics[height=1.7ex]{orcid-logo.png}}\textsuperscript{\,20}},
{E.\;Kashirin~\href{https://orcid.org/0000-0001-6062-7997}{\includegraphics[height=1.7ex]{orcid-logo.png}}\textsuperscript{\,20}},
{M.\;Kie{\l}bowicz~\href{https://orcid.org/0000-0002-4403-9201}{\includegraphics[height=1.7ex]{orcid-logo.png}}\textsuperscript{\,12}},
{V.A.\;Kireyeu~\href{https://orcid.org/0000-0002-5630-9264}{\includegraphics[height=1.7ex]{orcid-logo.png}}\textsuperscript{\,20}},
{R.\;Kolesnikov~\href{https://orcid.org/0009-0006-4224-1058}{\includegraphics[height=1.7ex]{orcid-logo.png}}\textsuperscript{\,20}},
{D.\;Kolev~\href{https://orcid.org/0000-0002-9203-4739}{\includegraphics[height=1.7ex]{orcid-logo.png}}\textsuperscript{\,2}},
{Y.\;Koshio\textsuperscript{\,8}},
{V.N.\;Kovalenko~\href{https://orcid.org/0000-0001-6012-6615}{\includegraphics[height=1.7ex]{orcid-logo.png}}\textsuperscript{\,20}},
{S.\;Kowalski~\href{https://orcid.org/0000-0001-9888-4008}{\includegraphics[height=1.7ex]{orcid-logo.png}}\textsuperscript{\,16}},
{B.\;Koz{\l}owski~\href{https://orcid.org/0000-0001-8442-2320}{\includegraphics[height=1.7ex]{orcid-logo.png}}\textsuperscript{\,19}},
{A.\;Krasnoperov~\href{https://orcid.org/0000-0002-1425-2861}{\includegraphics[height=1.7ex]{orcid-logo.png}}\textsuperscript{\,20}},
{W.\;Kucewicz~\href{https://orcid.org/0000-0002-2073-711X}{\includegraphics[height=1.7ex]{orcid-logo.png}}\textsuperscript{\,15}},
{M.\;Kuchowicz~\href{https://orcid.org/0000-0003-3174-585X}{\includegraphics[height=1.7ex]{orcid-logo.png}}\textsuperscript{\,18}},
{M.\;Kuich~\href{https://orcid.org/0000-0002-6507-8699}{\includegraphics[height=1.7ex]{orcid-logo.png}}\textsuperscript{\,17}},
{A.\;Kurepin~\href{https://orcid.org/0000-0002-1851-4136}{\includegraphics[height=1.7ex]{orcid-logo.png}}\textsuperscript{\,20}},
{A.\;L\'aszl\'o~\href{https://orcid.org/0000-0003-2712-6968}{\includegraphics[height=1.7ex]{orcid-logo.png}}\textsuperscript{\,5}},
{M.\;Lewicki~\href{https://orcid.org/0000-0002-8972-3066}{\includegraphics[height=1.7ex]{orcid-logo.png}}\textsuperscript{\,18}},
{G.\;Lykasov~\href{https://orcid.org/0000-0002-1544-6959}{\includegraphics[height=1.7ex]{orcid-logo.png}}\textsuperscript{\,20}},
{V.V.\;Lyubushkin~\href{https://orcid.org/0000-0003-0136-233X}{\includegraphics[height=1.7ex]{orcid-logo.png}}\textsuperscript{\,20}},
{M.\;Ma\'ckowiak-Paw{\l}owska~\href{https://orcid.org/0000-0003-3954-6329}{\includegraphics[height=1.7ex]{orcid-logo.png}}\textsuperscript{\,19}},
{Z.\;Majka~\href{https://orcid.org/0000-0003-3064-6577}{\includegraphics[height=1.7ex]{orcid-logo.png}}\textsuperscript{\,14}},
{A.\;Makhnev~\href{https://orcid.org/0009-0002-9745-1897}{\includegraphics[height=1.7ex]{orcid-logo.png}}\textsuperscript{\,20}},
{B.\;Maksiak~\href{https://orcid.org/0000-0002-7950-2307}{\includegraphics[height=1.7ex]{orcid-logo.png}}\textsuperscript{\,13}},
{A.I.\;Malakhov~\href{https://orcid.org/0000-0001-8569-8409}{\includegraphics[height=1.7ex]{orcid-logo.png}}\textsuperscript{\,20}},
{A.\;Marcinek~\href{https://orcid.org/0000-0001-9922-743X}{\includegraphics[height=1.7ex]{orcid-logo.png}}\textsuperscript{\,12}},
{A.D.\;Marino~\href{https://orcid.org/0000-0002-1709-538X}{\includegraphics[height=1.7ex]{orcid-logo.png}}\textsuperscript{\,24}},
{H.-J.\;Mathes~\href{https://orcid.org/0000-0002-0680-040X}{\includegraphics[height=1.7ex]{orcid-logo.png}}\textsuperscript{\,4}},
{T.\;Matulewicz~\href{https://orcid.org/0000-0003-2098-1216}{\includegraphics[height=1.7ex]{orcid-logo.png}}\textsuperscript{\,17}},
{V.\;Matveev~\href{https://orcid.org/0000-0002-2745-5908}{\includegraphics[height=1.7ex]{orcid-logo.png}}\textsuperscript{\,20}},
{G.L.\;Melkumov~\href{https://orcid.org/0009-0004-2074-6755}{\includegraphics[height=1.7ex]{orcid-logo.png}}\textsuperscript{\,20}},
{A.\;Merzlaya~\href{https://orcid.org/0000-0002-6553-2783}{\includegraphics[height=1.7ex]{orcid-logo.png}}\textsuperscript{\,10}},
{{\L}.\;Mik~\href{https://orcid.org/0000-0003-2712-6861}{\includegraphics[height=1.7ex]{orcid-logo.png}}\textsuperscript{\,15}},
{S.\;Morozov~\href{https://orcid.org/0000-0002-6748-7277}{\includegraphics[height=1.7ex]{orcid-logo.png}}\textsuperscript{\,20}},
{Y.\;Nagai~\href{https://orcid.org/0000-0002-1792-5005}{\includegraphics[height=1.7ex]{orcid-logo.png}}\textsuperscript{\,6}},
{T.\;Nakadaira~\href{https://orcid.org/0000-0003-4327-7598}{\includegraphics[height=1.7ex]{orcid-logo.png}}\textsuperscript{\,7}},
{M.\;Naskr\k{e}t~\href{https://orcid.org/0000-0002-5634-6639}{\includegraphics[height=1.7ex]{orcid-logo.png}}\textsuperscript{\,18}},
{S.\;Nishimori~\href{https://orcid.org/~0000-0002-1820-0938}{\includegraphics[height=1.7ex]{orcid-logo.png}}\textsuperscript{\,7}},
{A.\;Olivier~\href{https://orcid.org/0000-0003-4261-8303}{\includegraphics[height=1.7ex]{orcid-logo.png}}\textsuperscript{\,23}},
{V.\;Ozvenchuk~\href{https://orcid.org/0000-0002-7821-7109}{\includegraphics[height=1.7ex]{orcid-logo.png}}\textsuperscript{\,12}},
{O.\;Panova~\href{https://orcid.org/0000-0001-5039-7788}{\includegraphics[height=1.7ex]{orcid-logo.png}}\textsuperscript{\,11}},
{V.\;Paolone~\href{https://orcid.org/0000-0003-2162-0957}{\includegraphics[height=1.7ex]{orcid-logo.png}}\textsuperscript{\,26}},
{O.\;Petukhov~\href{https://orcid.org/0000-0002-8872-8324}{\includegraphics[height=1.7ex]{orcid-logo.png}}\textsuperscript{\,20}},
{I.\;Pidhurskyi~\href{https://orcid.org/0000-0001-9916-9436}{\includegraphics[height=1.7ex]{orcid-logo.png}}\textsuperscript{\,11}},
{R.\;P{\l}aneta~\href{https://orcid.org/0000-0001-8007-8577}{\includegraphics[height=1.7ex]{orcid-logo.png}}\textsuperscript{\,14}},
{P.\;Podlaski~\href{https://orcid.org/0000-0002-0232-9841}{\includegraphics[height=1.7ex]{orcid-logo.png}}\textsuperscript{\,17}},
{B.A.\;Popov~\href{https://orcid.org/0000-0001-5416-9301}{\includegraphics[height=1.7ex]{orcid-logo.png}}\textsuperscript{\,20,3}},
{B.\;P\'orfy~\href{https://orcid.org/0000-0001-5724-9737}{\includegraphics[height=1.7ex]{orcid-logo.png}}\textsuperscript{\,5,6}},
{D.S.\;Prokhorova~\href{https://orcid.org/0000-0003-3726-9196}{\includegraphics[height=1.7ex]{orcid-logo.png}}\textsuperscript{\,20}},
{D.\;Pszczel~\href{https://orcid.org/0000-0002-4697-6688}{\includegraphics[height=1.7ex]{orcid-logo.png}}\textsuperscript{\,13}},
{S.\;Pu{\l}awski~\href{https://orcid.org/0000-0003-1982-2787}{\includegraphics[height=1.7ex]{orcid-logo.png}}\textsuperscript{\,16}},
{J.\;Puzovi\'c\textsuperscript{\,21}\textsuperscript{\dag}},
{R.\;Renfordt~\href{https://orcid.org/0000-0002-5633-104X}{\includegraphics[height=1.7ex]{orcid-logo.png}}\textsuperscript{\,16}},
{L.\;Ren~\href{https://orcid.org/0000-0003-1709-7673}{\includegraphics[height=1.7ex]{orcid-logo.png}}\textsuperscript{\,24}},
{V.Z.\;Reyna~Ortiz~\href{https://orcid.org/0000-0002-7026-8198}{\includegraphics[height=1.7ex]{orcid-logo.png}}\textsuperscript{\,11}},
{D.\;R\"ohrich\textsuperscript{\,9}},
{E.\;Rondio~\href{https://orcid.org/0000-0002-2607-4820}{\includegraphics[height=1.7ex]{orcid-logo.png}}\textsuperscript{\,13}},
{M.\;Roth~\href{https://orcid.org/0000-0003-1281-4477}{\includegraphics[height=1.7ex]{orcid-logo.png}}\textsuperscript{\,4}},
{{\L}.\;Rozp{\l}ochowski~\href{https://orcid.org/0000-0003-3680-6738}{\includegraphics[height=1.7ex]{orcid-logo.png}}\textsuperscript{\,12}},
{B.T.\;Rumberger~\href{https://orcid.org/0000-0002-4867-945X}{\includegraphics[height=1.7ex]{orcid-logo.png}}\textsuperscript{\,24}},
{M.\;Rumyantsev~\href{https://orcid.org/0000-0001-8233-2030}{\includegraphics[height=1.7ex]{orcid-logo.png}}\textsuperscript{\,20}},
{A.\;Rustamov~\href{https://orcid.org/0000-0001-8678-6400}{\includegraphics[height=1.7ex]{orcid-logo.png}}\textsuperscript{\,1}},
{M.\;Rybczynski~\href{https://orcid.org/0000-0002-3638-3766}{\includegraphics[height=1.7ex]{orcid-logo.png}}\textsuperscript{\,11}},
{A.\;Rybicki~\href{https://orcid.org/0000-0003-3076-0505}{\includegraphics[height=1.7ex]{orcid-logo.png}}\textsuperscript{\,12}},
{D.\;Rybka\textsuperscript{\,13}},
{K.\;Sakashita~\href{https://orcid.org/0000-0003-2602-7837}{\includegraphics[height=1.7ex]{orcid-logo.png}}\textsuperscript{\,7}},
{K.\;Schmidt~\href{https://orcid.org/0000-0002-0903-5790}{\includegraphics[height=1.7ex]{orcid-logo.png}}\textsuperscript{\,16}},
{A.Yu.\;Seryakov~\href{https://orcid.org/0000-0002-5759-5485}{\includegraphics[height=1.7ex]{orcid-logo.png}}\textsuperscript{\,20}},
{P.\;Seyboth~\href{https://orcid.org/0000-0002-4821-6105}{\includegraphics[height=1.7ex]{orcid-logo.png}}\textsuperscript{\,11}},
{U.A.\;Shah~\href{https://orcid.org/0000-0002-9315-1304}{\includegraphics[height=1.7ex]{orcid-logo.png}}\textsuperscript{\,11}},
{Y.\;Shiraishi\textsuperscript{\,8}},
{A.\;Shukla~\href{https://orcid.org/0000-0003-3839-7229}{\includegraphics[height=1.7ex]{orcid-logo.png}}\textsuperscript{\,25}},
{M.\;S{\l}odkowski~\href{https://orcid.org/0000-0003-0463-2753}{\includegraphics[height=1.7ex]{orcid-logo.png}}\textsuperscript{\,19}},
{P.\;Staszel~\href{https://orcid.org/0000-0003-4002-1626}{\includegraphics[height=1.7ex]{orcid-logo.png}}\textsuperscript{\,14}},
{G.\;Stefanek~\href{https://orcid.org/0000-0001-6656-9177}{\includegraphics[height=1.7ex]{orcid-logo.png}}\textsuperscript{\,11}},
{J.\;Stepaniak~\href{https://orcid.org/0000-0003-2064-9870}{\includegraphics[height=1.7ex]{orcid-logo.png}}\textsuperscript{\,13}},
{M.\;Strikhanov\textsuperscript{\,20}},
{{\L}.\;\'Swiderski~\href{https://orcid.org/0000-0001-5857-2085}{\includegraphics[height=1.7ex]{orcid-logo.png}}\textsuperscript{\,13}},
{J.\;Szewi\'nski~\href{https://orcid.org/0000-0003-2981-9303}{\includegraphics[height=1.7ex]{orcid-logo.png}}\textsuperscript{\,13}},
{R.\;Szukiewicz~\href{https://orcid.org/0000-0002-1291-4040}{\includegraphics[height=1.7ex]{orcid-logo.png}}\textsuperscript{\,18}},
{A.\;Taranenko~\href{https://orcid.org/0000-0003-1737-4474}{\includegraphics[height=1.7ex]{orcid-logo.png}}\textsuperscript{\,20}},
{A.\;Tefelska~\href{https://orcid.org/0000-0002-6069-4273}{\includegraphics[height=1.7ex]{orcid-logo.png}}\textsuperscript{\,19}},
{D.\;Tefelski~\href{https://orcid.org/0000-0003-0802-2290}{\includegraphics[height=1.7ex]{orcid-logo.png}}\textsuperscript{\,19}},
{V.\;Tereshchenko\textsuperscript{\,20}},
{R.\;Tsenov~\href{https://orcid.org/0000-0002-1330-8640}{\includegraphics[height=1.7ex]{orcid-logo.png}}\textsuperscript{\,2}},
{L.\;Turko~\href{https://orcid.org/0000-0002-5474-8650}{\includegraphics[height=1.7ex]{orcid-logo.png}}\textsuperscript{\,18}},
{T.S.\;Tveter~\href{https://orcid.org/0009-0003-7140-8644}{\includegraphics[height=1.7ex]{orcid-logo.png}}\textsuperscript{\,10}},
{M.\;Unger~\href{https://orcid.org/0000-0002-7651-0272~}{\includegraphics[height=1.7ex]{orcid-logo.png}}\textsuperscript{\,4}},
{M.\;Urbaniak~\href{https://orcid.org/0000-0002-9768-030X}{\includegraphics[height=1.7ex]{orcid-logo.png}}\textsuperscript{\,16}},
{F.F.\;Valiev~\href{https://orcid.org/0000-0001-5130-5603}{\includegraphics[height=1.7ex]{orcid-logo.png}}\textsuperscript{\,20}},
{D.\;Veberi\v{c}~\href{https://orcid.org/0000-0003-2683-1526}{\includegraphics[height=1.7ex]{orcid-logo.png}}\textsuperscript{\,4}},
{V.V.\;Vechernin~\href{https://orcid.org/0000-0003-1458-8055}{\includegraphics[height=1.7ex]{orcid-logo.png}}\textsuperscript{\,20}},
{O.\;Vitiuk~\href{https://orcid.org/0000-0002-9744-3937}{\includegraphics[height=1.7ex]{orcid-logo.png}}\textsuperscript{\,18}},
{V.\;Volkov~\href{https://orcid.org/0000-0002-4785-7517}{\includegraphics[height=1.7ex]{orcid-logo.png}}\textsuperscript{\,20}},
{A.\;Wickremasinghe~\href{https://orcid.org/0000-0002-5325-0455}{\includegraphics[height=1.7ex]{orcid-logo.png}}\textsuperscript{\,22}},
{K.\;Witek~\href{https://orcid.org/0009-0004-6699-1895}{\includegraphics[height=1.7ex]{orcid-logo.png}}\textsuperscript{\,15}},
{K.\;W\'ojcik~\href{https://orcid.org/0000-0002-8315-9281}{\includegraphics[height=1.7ex]{orcid-logo.png}}\textsuperscript{\,16}},
{O.\;Wyszy\'nski~\href{https://orcid.org/0000-0002-6652-0450}{\includegraphics[height=1.7ex]{orcid-logo.png}}\textsuperscript{\,11}},
{A.\;Zaitsev~\href{https://orcid.org/0000-0003-4711-9925}{\includegraphics[height=1.7ex]{orcid-logo.png}}\textsuperscript{\,20}},
{E.\;Zherebtsova~\href{https://orcid.org/0000-0002-1364-0969}{\includegraphics[height=1.7ex]{orcid-logo.png}}\textsuperscript{\,18}},
{E.D.\;Zimmerman~\href{https://orcid.org/0000-0002-6394-6659}{\includegraphics[height=1.7ex]{orcid-logo.png}}\textsuperscript{\,24}},
{A.\;Zviagina~\href{https://orcid.org/0009-0007-5211-6493}{\includegraphics[height=1.7ex]{orcid-logo.png}}\textsuperscript{\,20}}, and
{R.\;Zwaska~\href{https://orcid.org/0000-0002-4889-5988}{\includegraphics[height=1.7ex]{orcid-logo.png}}\textsuperscript{\,22}}
\\\rule{2cm}{.5pt}\\[-.5ex]\textit{\textsuperscript{\dag} \footnotesize deceased}

\end{sloppypar}

\noindent
\textsuperscript{1}~National Nuclear Research Center, Baku, Azerbaijan\\
\textsuperscript{2}~Faculty of Physics, University of Sofia, Sofia, Bulgaria\\
\textsuperscript{3}~LPNHE, Sorbonne University, CNRS/IN2P3, Paris, France\\
\textsuperscript{4}~Karlsruhe Institute of Technology, Karlsruhe, Germany\\
\textsuperscript{5}~HUN-REN Wigner Research Centre for Physics, Budapest, Hungary\\
\textsuperscript{6}~E\"otv\"os Lor\'and University, Budapest, Hungary\\
\textsuperscript{7}~Institute for Particle and Nuclear Studies, Tsukuba, Japan\\
\textsuperscript{8}~Okayama University, Japan\\
\textsuperscript{9}~University of Bergen, Bergen, Norway\\
\textsuperscript{10}~University of Oslo, Oslo, Norway\\
\textsuperscript{11}~Jan Kochanowski University, Kielce, Poland\\
\textsuperscript{12}~Institute of Nuclear Physics, Polish Academy of Sciences, Cracow, Poland\\
\textsuperscript{13}~National Centre for Nuclear Research, Warsaw, Poland\\
\textsuperscript{14}~Jagiellonian University, Cracow, Poland\\
\textsuperscript{15}~AGH - University of Science and Technology, Cracow, Poland\\
\textsuperscript{16}~University of Silesia, Katowice, Poland\\
\textsuperscript{17}~University of Warsaw, Warsaw, Poland\\
\textsuperscript{18}~University of Wroc{\l}aw,  Wroc{\l}aw, Poland\\
\textsuperscript{19}~Warsaw University of Technology, Warsaw, Poland\\
\textsuperscript{20}~Affiliated with an institution covered by a cooperation agreement with CERN\\
\textsuperscript{21}~University of Belgrade, Belgrade, Serbia\\
\textsuperscript{22}~Fermilab, Batavia, USA\\
\textsuperscript{23}~University of Notre Dame, Notre Dame, USA\\
\textsuperscript{24}~University of Colorado, Boulder, USA\\
\textsuperscript{25}~University of Hawaii at Manoa, Honolulu, USA\\
\textsuperscript{26}~University of Pittsburgh, Pittsburgh, USA\\

\end{document}